\documentclass{nature}
\usepackage{graphicx}
\usepackage[nointegrals]{wasysym}
\usepackage{rotating}
\makeatletter
\let\saved@includegraphics\includegraphics
\AtBeginDocument{\let\includegraphics\saved@includegraphics}
\renewenvironment*{figure}{\@float{figure}}{\end@float}
\makeatother

\usepackage[margin=0.5in]{geometry}
\usepackage[utf8]{inputenc}
\usepackage{textcomp}
\usepackage[version=4]{mhchem}
\usepackage{array}
\newcolumntype{L}{>{\flushleft\arraybackslash}m{7cm}}
\newcolumntype{Q}{>{\flushleft\arraybackslash}m{4cm}}
\newcolumntype{M}{>{\flushleft\arraybackslash}m{2cm}}
\newcolumntype{D}{>{\flushleft\arraybackslash}m{2em}}
\newcolumntype{O}{>{\flushleft\arraybackslash}m{12cm}}

\newcommand{\beginextended}{%
        \setcounter{table}{0}
        \renewcommand{\thetable}{E\arabic{table}}%
        \setcounter{figure}{0}
        \renewcommand{\thefigure}{E\arabic{figure}}%
        }

\newcommand{\beginsupplement}{%
        \setcounter{table}{0}
        \renewcommand{\thetable}{S\arabic{table}}%
        \setcounter{figure}{0}
        \renewcommand{\thefigure}{S\arabic{figure}}%
        }

\usepackage{lineno}

\title{A nitrogen-rich atmosphere on ancient Mars consistent with isotopic evolution models}

\author{Renyu Hu$^{1,2}$ and Trent B. Thomas$^{1,3}$}

\begin{document}

\maketitle

\begin{affiliations}
 \item Jet Propulsion Laboratory, California Institute of Technology, Pasadena, CA 91109, USA, email: renyu.hu@jpl.nasa.gov
 \item Division of Geological and Planetary Sciences, California Institute of Technology, Pasadena, CA 91125, USA
 \item Department of Earth and Space Sciences and Astrobiology Program, University of Washington, Seattle, WA 98195, USA
\end{affiliations}


\begin{abstract}
The ratio of nitrogen isotopes in the Martian atmosphere is a key constraint on the planet's atmospheric evolution. However, enrichment of the heavy isotope expected due to atmospheric loss from sputtering and photochemical processes is greater than measurements. A massive, multi-bar early \ce{CO2}-dominated atmosphere and recent volcanic outgassing have been proposed to explain this discrepancy, and many previous models have assumed atmospheric nitrogen rapidly reached a steady state where loss to space balanced volcanic outgassing. Here we show using time-dependent models that the abundance and isotopic composition of nitrogen in the Martian atmosphere can be explained by a family of evolutionary scenarios in which the initial partial pressure of nitrogen is sufficiently high that a steady state is not reached and nitrogen levels gradually decline to present-day values over 4 billion years. Our solutions do not require a multi-bar early \ce{CO2} atmosphere and are consistent with volcanic outgassing indicated by both geologic mapping and the atmospheric $^{36}$Ar/$^{38}$Ar ratio. Monte Carlo simulations that include these scenarios estimate that the partial pressure of \ce{N2} was 60 -- 740 mbar (90\% confidence, with a median value of 310 mbar) at 3.8 billion years ago when the valley networks formed. We suggest that such a high nitrogen partial pressure could have contributed substantially to warming on early Mars.
\end{abstract}

We have constructed a multi-functional model for the evolution of the nitrogen's abundance and isotopic composition in Mars's atmosphere and regolith, with volcanic outgassing, escape to space, and nitrate deposition as sources and sinks (Methods and Fig.~\ref{fig:boxmod}). The model starts at 3.8 Ga, i.e., after the last major impact ($\sim3.9$ Ga\cite{fassett2011sequence,robbins2013large}), and from a $\delta^{15}$N value of the mantle component measured in the Martian meteorite ALH~84001\cite{mathew2001early}. $\delta^{15}$N is defined as the relative enhancement of the ratio $^{15}$N/$^{14}$N with respect to a reference standard (Earth's atmosphere). During the modeled period, the escape processes include photochemical escape, sputtering, and ion escape. We seek to reproduce the present-day size and isotopic composition of the \ce{N2} reservoir\cite{Wong2013}. The model has seven parameters, listed in Table~\ref{table:priors}, to capture the uncertainty in the rates of escape, volcanic outgassing, and nitrate deposition, as well as the extent of diffusive fractionation between the bulk atmosphere and the exobase, where escape takes place.

Compared with previous evolutionary models of Mars nitrogen\cite{mcelroy1976isotopic,McElroy1976,Fox1983,Fox1993,Jakosky1994,Zent1994,FoxHac1997,Kurokawa2018} (Supplementary Information A), new aspects of this model include: (1) a revised photochemical escape rate and fractionation factor; (2) inclusion of nitrate deposition guided by its recent discovery\cite{stern2015,sutter2017evolved}; (3) constraints of the volcanic outgassing history from argon isotopes\cite{Slipski2016}; and (4) coupling with the \ce{CO2} evolution histories constrained by carbon isotopes\cite{Hu2015}. We have made a substantial revision to the escape rate of the photodissociation of \ce{N2} and have calculated the fractionation factor in this process, using the photochemical isotope effect method\cite{Hu2015} and the recent experimental results on the \ce{N2} photodissociation channels\cite{song2016quantum} (Supplementary Information B). Because the escape rate of nitrogen is also proportional to the mixing ratio of \ce{N2} with respect to \ce{CO2} when nitrogen is a relatively minor species in the atmosphere, the nitrogen evolution model must be coupled with evolutionary scenarios of \ce{CO2}. We have applied representative \ce{CO2} evolution scenarios\cite{Hu2015}, where the initial pressure ranges in 0.25 -- 1.8 bars (Fig.~\ref{fig:CO2}), and adopted the \ce{CO2} scenario with an initial partial pressure of 1.0 bar as the default because it may be more consistent with the comprehensive extrapolation from Mars Atmosphere and Volatile
Evolution (MAVEN) measurements\cite{jakosky2018loss,jakosky2019co2} and required to cause water
ice melting on early Mars's surface\cite{forget20133d,von2013n2}.

\begin{sidewaystable}[!htbp]
\centering
\begin{tabular}{llll} 
 \hline\hline
 Parameter & Symbol & Unconstrained MCMC & Constrained MCMC \\ 
 \hline\hline
Power-law index of photochemical loss & $a$ & 0.5 - 3.0 & 0.5 - 3.0  \\
Multiplier of photochemical loss & $f_{\rm pr}$ & 0.001 - 100 & 0.001 - 2 \\
Multiplier of sputtering loss & $f_{\rm sp}$ & 0.001 - 100 & 0.001 - 2 \\
Multiplier of volcanic outgassing & $f_{\rm og}$ & 0.001 - 100 & 0.001 - 2 \\
Diffusion parameter (km/K) & $\Delta z/T$ & 0.2 - 0.5 & 0.2 - 0.5 \\
Depth of nitrate deposition (m) & $d$ & 0.001 - 1000 & 0.001 - 1000 \\
Partial pressure of \ce{N2} at 3.8 Ga (mbar) & $P_{\rm 3.8Ga}$ & 0.001 - 1000 & 0.001 - 1000 \\
 \hline
\end{tabular}
\caption{Parameters of the nitrogen evolution model. 
The photochemical escape rate is extrapolated from the current-epoch estimate using a power-law scaling to the solar EUV flux. Multipliers are then applied to the baseline models of photochemical loss, sputtering loss, and volcanic outgassing to allow exploration of the strengths of these processes. The constrained MCMCs apply plausible upper limits of the multipliers, and the unconstrained MCMCs explore the multipliers in essentially unlimited ranges.
The diffusion parameter is the difference between the altitude of the homopause and the exobase divided by the temperature in this interval, and this parameter has been measured by MAVEN\cite{jakosky2017}.
Finally, the depth of nitrate deposition describes how much nitrate has formed and been sequestered in the crust.
Full definitions of these parameters are provided in Methods.}
\label{table:priors}
\end{sidewaystable}

\subsection{Bifurcation.} 
\begin{figure}[!htbp]
\centering
\includegraphics[width=0.8\textwidth]{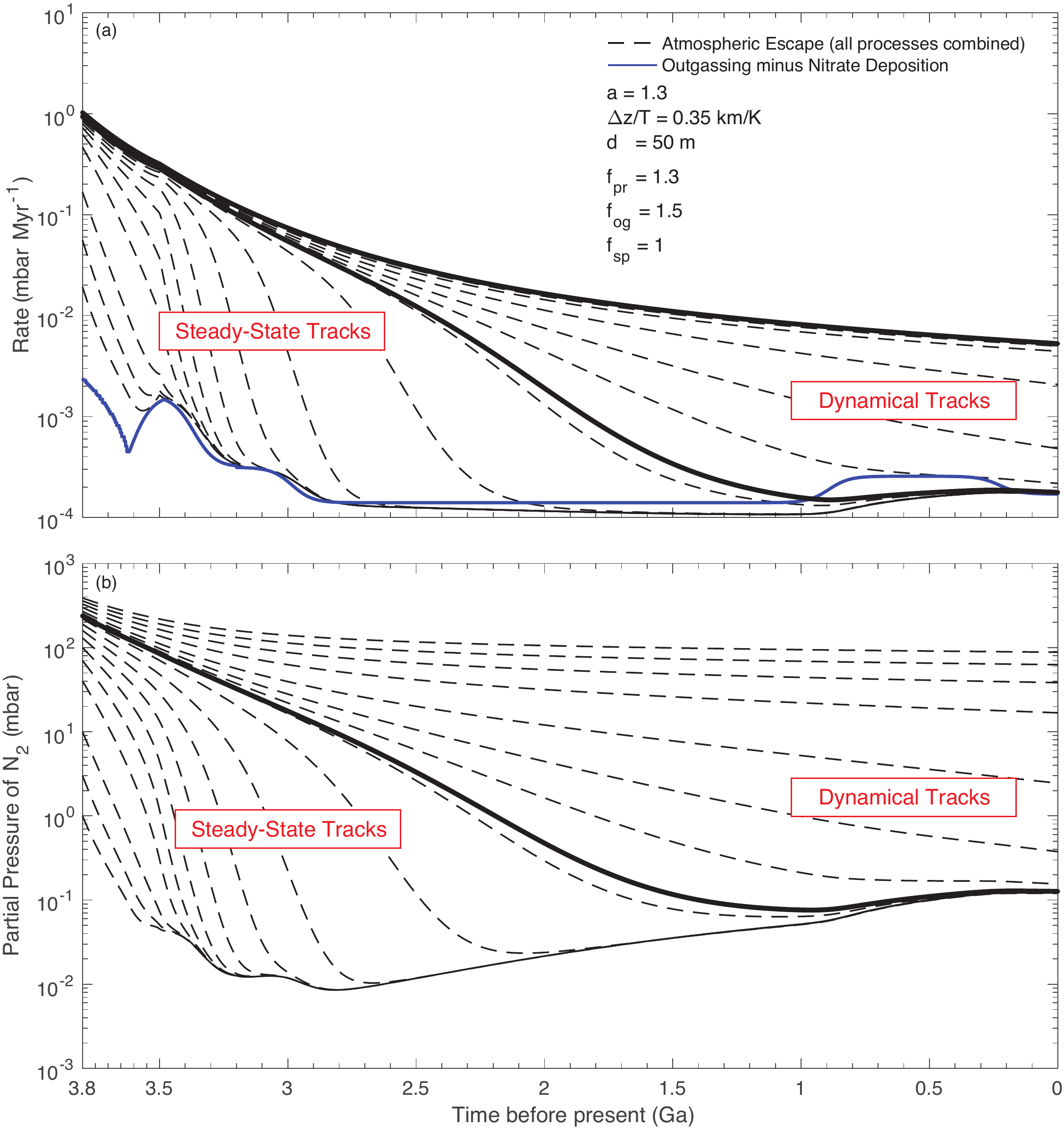}
\caption{Bifurcation in the evolution of nitrogen between the steady-state scenarios and the dynamical scenarios. The panel (a) and (b) show the change rates and the partial pressure of \ce{N2} for evolutionary tracks with the same set of parameters but different initial pressures. When the initial pressure is greater than a critical value, the steady state cannot be reached by the end and a new set of dynamical scenarios emerges. The transitional track from this critical initial pressure is shown in a thick line, and it separates the steady-state and the dynamical scenarios.
}
\label{fig:bifur}
\end{figure}

We discovered a bifurcation between the evolutionary tracks starting from a low \ce{N2} partial pressure and those from a high one. The low-start solutions feature the ``steady state,'' where atmospheric loss balances outgassing and nitrate deposition, while the high-start solutions never reach the steady state (Fig.~\ref{fig:bifur}). When the initial \ce{N2} partial pressure is low, the rate of atmospheric escape quickly converges with the combined rate of outgassing and nitrate deposition, i.e., the steady state. Over a wide range of the initial pressure, it takes a few hundred million years to reach the steady state. The time it takes, however, does increases for a higher initial pressure. When the initial pressure is high enough (e.g., a few 100 mbar or more), the steady state is never reached within the age of the Solar System, and a different family of evolution tracks emerges. We call this new family `dynamical.' The dynamical tracks have a protracted descent of the partial pressure of \ce{N2} during evolution rather than a speedy descent to the steady state. While the steady-state tracks converge to the same final partial pressure, the dynamical tracks lead to different final partial pressures depending on the initial value. For the same set of parameters, the final pressures of the dynamical tracks are always higher than the final pressure of the steady-state tracks (Fig.~\ref{fig:bifur}).

The bifurcation can be understood with the help of the following idealized and analytical model. Let $X$ be the size of the nitrogen reservoir, $V$ be the addition rate to this reservoir by volcanic outgassing (discounted by the rate of nitrate deposition, if any), and $E$ be the escape rate. When the size of the nitrogen reservoir is small compared with the carbon reservoir, the escape is proportional to its size (Methods), and at the limit of a constant background \ce{CO2} pressure, we can write $E=eX$, where $e$ is a coefficient. The equation for $X$ is then
\begin{equation}
    \frac{dX}{dt}=V-eX.
\end{equation}
For a constant $V$ and $e$, the equation has following solution
\begin{equation}
    X=\frac{V}{e} + \bigg(X_0-\frac{V}{e} \bigg)\exp(-et),\label{eq:evosim}
\end{equation}
where $X_0$ is $X$ at $t=0$. Therefore, the solution converges to the steady state with a timescale of $1/e$. Note that the solution in Eq. (\ref{eq:evosim}), despite its idealized nature, is quite similar to the segment of a given steady-state solution that rapidly approaches the steady state (Fig.~\ref{fig:bifur}). Currently the $1/e$ timescale for sputtering loss is $\sim4000$ Myr\cite{leblanc2002}, and that for photochemical loss is $\sim1000$ Myr\cite{Fox1993}. The corresponding timescales were $\sim2$ Myr and $\sim70$ Myr (for a power-law index of 1.5) at 3.5 Ga when the solar extreme ultraviolet (EUV) flux was approximately 6 times the current value (see Methods). These estimates tell us that (1) it is possible to reach the steady state in the early evolution and it becomes harder as time goes by; and (2) sputtering loss is the main process that drives the system to the steady state in the early epoch (because of its small $1/e$ timescale). 

Then, why do some evolution tracks never reach the steady state? This is because the sputtering rate is no longer proportional to the size of the nitrogen reservoir when the atmosphere is moderately nitrogen-rich (see Eq.~\ref{eq:sputdep} in Methods). The sputtering rate decouples when the size of the nitrogen reservoir approaches the size of the carbon reservoir, corrected by the diffusive separation factor between \ce{N2} and \ce{CO2}. This factor is $\sim8$ for typical upper-atmosphere conditions\cite{jakosky2017}, implying that \ce{N2} cannot be considered a minor component for determining the sputtering rate when it is approximately 10\% or more in the bulk atmosphere. Because of this decoupling, $E$ can no longer be written as $eX$ and the steady state no longer exists. The bifurcation can thus be understood as such: when the initial nitrogen pressure is small, the system can quickly reach the steady state following the solution in Eq. (\ref{eq:evosim}); when the initial nitrogen pressure is large, the atmosphere must first lose nitrogen relative to carbon, on a linear rather than exponential time dependency, and by the time it starts to converge to the steady state, the $1/e$ timescale is already large, and then the steady state can never be reached within the age of the Solar System.

\begin{figure}[!htbp]
\centering
\includegraphics[width=0.8\textwidth]{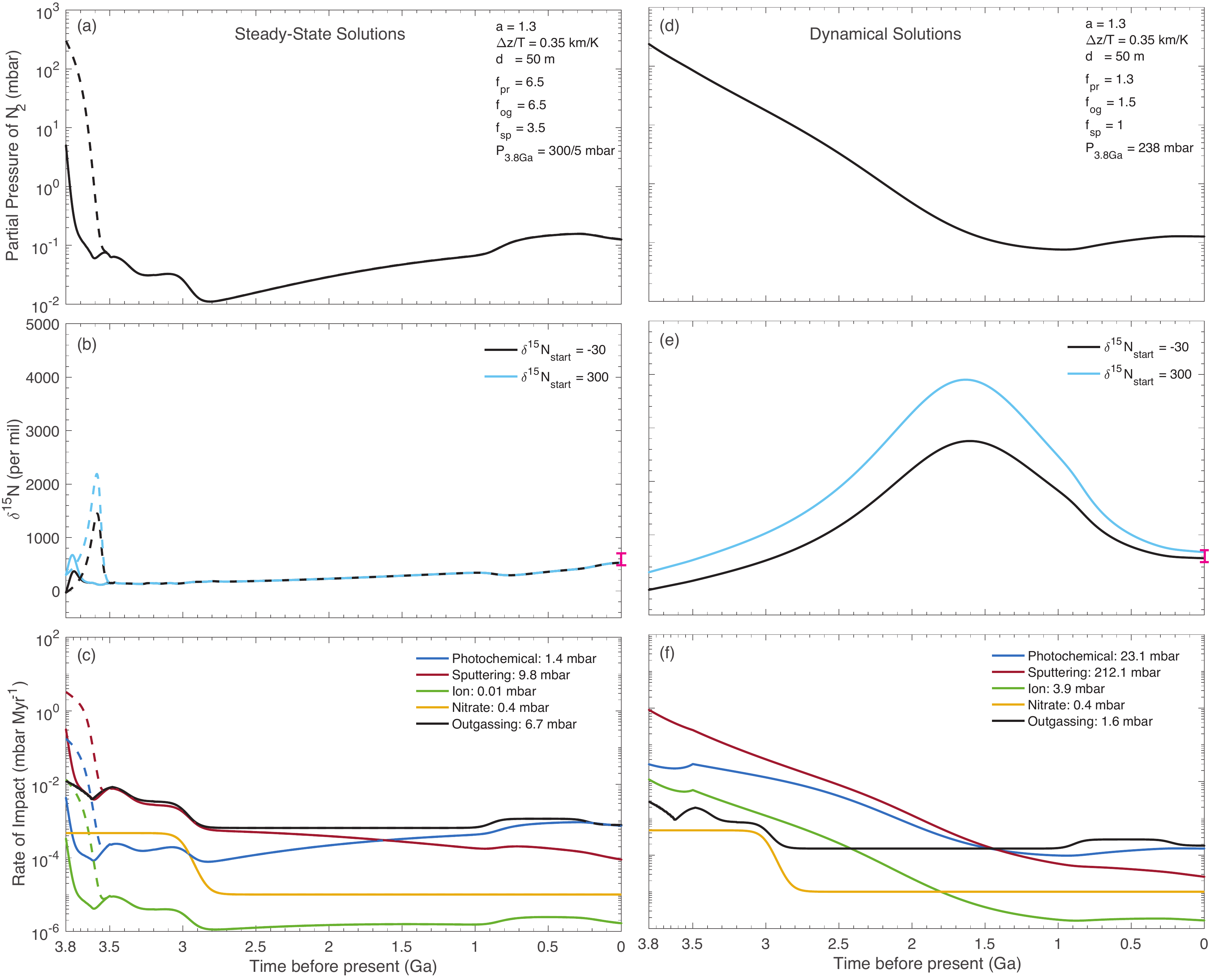}
\caption{Example of the steady-state and dynamical solutions consistent with the size and isotopic composition of Mars’s nitrogen reservoir. These models adopt the \ce{CO2} evolutionary scenario 3 with the initial partial pressure of 1.0 bar (Fig.~\ref{fig:CO2}). Two steady-state solutions with different initial pressures are shown with solid and dashed lines. In panels (b) and (e), the solutions starting with an elevated initial $\delta^{15}$N of $300\permil$\cite{Kurokawa2018} are shown in blue for comparison, and the error bar in purple is the Curiosity measurement\cite{Wong2013}. 
In panels (c) and (f), the labels show the total mass added or removed by each process for the solutions shown in the solid lines.
The steady-state and dynamical solutions require distinct parameters to match the present-day abundance and isotopic composition.
}
\label{fig:example}
\end{figure}

\subsection{Steady-State and Dynamical Solutions.} The present-day size and isotopic composition of the nitrogen reservoir can be matched by either the steady-state or the dynamical evolution tracks by adjusting the parameters for the rates of outgassing and escape. Fig.~\ref{fig:example} shows example steady-state and dynamical solutions. In these solutions, sputtering removes most of the mass in total, while photochemical escape takes over as the dominant mass-loss mechanism around $1\sim2$ Ga before present. In the dynamical solution, the initial partial pressure is 238 mbar, and sputtering removes 212 mbar and photochemical escape removes 23 mbar. Nitrate deposition, which can be comparable to volcanic outgassing in rates $>3$ Ga before present, is not a dominant sink in the more recent history. Outgassing is an important process in the recent history, and as a result, specific implementation of the outgassing baseline model (Fig.~\ref{fig:crustal}) can affect the evolutionary outcome.

For the steady-state solutions shown in Fig.~\ref{fig:example}, increasing the initial partial pressure to a critical value of $\sim1.2$ bars does not change the evolutionary outcome, but an even higher initial pressure would lead to bifurcation and a dynamical track. That track is no longer a solution because the final pressure would be too high. For the dynamical solution shown in Fig.~\ref{fig:example}, which uses smaller rate multipliers ($f_{\rm sp}$ and $f_{\rm pr}$) compared to the steady-state solutions, an initial partial pressure of $\sim240$ mbar already causes the dynamical evolution. Further decreasing the initial partial pressure would result in a steady-state track in this case, but that track is not a solution either because the final $\delta^{15}$N would be substantially smaller than the observed. From these examples, we make the following three observations. (1) The insensitivity to the initial pressure of the steady-state solutions is consistent with the understanding in the past, that the memory of the initial state is `lost' for the nitrogen evolution\cite{Kurokawa2018}. However, this understanding only applies to the scenarios where the initial partial pressure is less than a critical value at which the bifurcation occurs. (2) The critical pressure increases for higher escape rates or greater photochemical loss and sputtering multipliers ($f_{\rm pr}$ and $f_{\rm sp}$). (3) The dynamical solutions are typically the evolution tracks that are quite close to the transitional track between the steady-state and the dynamical tracks. This is not surprising because the final partial pressure of the dynamical tracks quickly diverge to large values, which would be inconsistent with the present-day Mars (Fig.~\ref{fig:bifur}). As a result of this adjacency to the steady-state solutions, the final isotopic composition only moderately depends on the initial value (Fig.~\ref{fig:example}). Because of all these features, if the dynamical solutions indeed apply to Mars, the current isotopic composition can be used to infer the partial pressure of nitrogen in the past.

\subsection{A Nitrogen-rich Early Martian Atmosphere.} 
\begin{figure}[!htbp]
\centering
\includegraphics[width=0.8\textwidth]{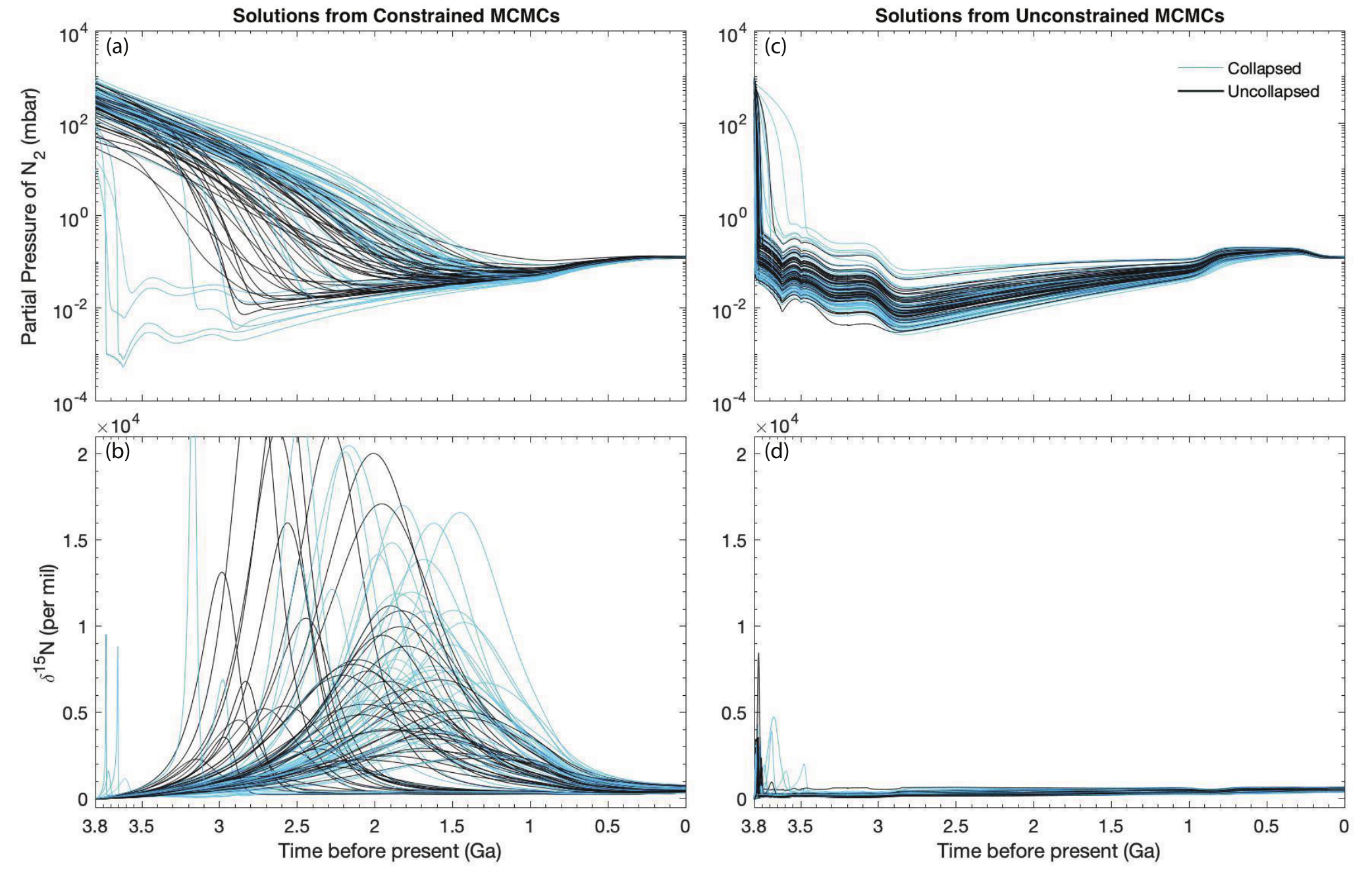}
\caption{Randomly selected solutions from the constrained and the unconstrained MCMCs. These models use the \ce{CO2} evolutionary scenario 3 with the initial partial pressure of 1.0 bar or the collapsed atmosphere scenario or the collapsed atmosphere scenario (Fig.~\ref{fig:CO2}). Each line represents an evolution path that meets the present-day partial pressure and isotopic composition of \ce{N2}. Most solutions from the unconstrained MCMCs are solutions that quickly converge to the steady states, and most solutions from the constrained MCMCs are dynamical solutions, or solutions that reach the steady states late in the evolution.
}
\label{fig:sample}
\end{figure}

We have systematically explored the potential evolutionary paths of Mars nitrogen with the Markov-Chain Monte Carlo (MCMC) method to fit the present-day partial pressure and isotopic composition. Using the parameters listed in Table~\ref{table:priors}, we have run `unconstrained MCMC' where the rate multipliers are allowed to vary in wide ranges, and also `constrained MCMC' where the outgassing multiplier is only allowed to vary within the upper limit derived from argon isotopes\cite{Slipski2016} and the escape rates are allowed to vary up to twice the baseline models\cite{Fox1993,leblanc2002}. Because the baseline volcanism model is consistent with the one adopted in the argon isotope study (Fig.~\ref{fig:crustal}), the argon-based constraint is applicable here to the extent that Mars's mantle has a similar N/Ar ratio as Earth. The constrained MCMCs can thus be understood as exploring the scenarios permitted by the argon isotopes and the current understanding of nonthermal escape from Mars. The posterior distributions of parameters are shown in Figs.~\ref{fig:post_100}-\ref{fig:post_sp_multi}, and solutions randomly selected from the unconstrained and constrained MCMCs are shown in Fig.~\ref{fig:sample}.

When there is essentially no limit on the rate multipliers, the steady-state solutions can be found with large escape and outgassing rates (Supplementary Information E). When we impose the upper limits on the rate multipliers (i.e., the constrained MCMCs), the escape rates are smaller, and thus the solutions take a longer time to reach the corresponding steady states, or cannot reach the steady states at all (Fig.~\ref{fig:sample}). In this case, the initial partial pressure of \ce{N2} is constrained despite all other model parameters (Figs.~\ref{fig:post_2} and \ref{fig:post_sp_multi}) and is not sensitive to the specific implementation of the baseline outgassing model, the initial $\delta^{15}$N value, or potential variation of the $\Delta z/T$ parameter over the course of evolution (Supplementary Information F). The spread in the posterior distribution of the initial pressure mainly comes from the uncertainty of the sputtering rate as sputtering is the dominant mass removal mechanism. Fig.~\ref{fig:post_sp_multi} shows that the posterior distribution becomes very narrow when the sputtering multiplier is fixed and moves to higher values when the multiplier increases.

For the sputtering rate within a factor of 2 from the current estimate\cite{leblanc2002}, the evolution model constrains the partial pressure of nitrogen at 3.8 Ga to 60 -- 740 mbar (90\% confidence), with a median value of 310 mbar (Fig.~\ref{fig:post_2}). The median value increases to 370 mbar when the atmosphere collapse is enforced. While we caution against placing too much emphasis on these specific values, the MCMCs clearly indicate a nitrogen-rich early Martian atmosphere. The ability to constrain the early abundance of \ce{N2} fundamentally comes from the solutions being (close to) dynamical when the realistic constraints from argon isotopes and nonthermal escape models are imposed. We thus suggest that the dynamical solutions or the solutions that reach the steady states late in the evolution may better represent the evolution of nitrogen on Mars. 

The dynamical solutions imply an \ce{N2}-rich atmosphere on early Mars, for which the escape rate of carbon would be reduced compared to the rate for a \ce{CO2}-dominated atmosphere. We thus present models of the \ce{CO2}-\ce{N2} evolution that include this feedback in Supplementary Information G. In essence, self-consistent solutions can be found with small changes of the input parameters, and the character of the steady-state and dynamical solutions remains unchanged. While the high initial \ce{N2} in the dynamical solutions may suppress the escape of carbon, the amount of early carbonate deposition is poorly constrained\cite{Hu2015} and so is the initial partial pressure of \ce{CO2}. Figs.~\ref{fig:post_100} and \ref{fig:post_2} show that the nitrogen evolution models and the associated constraints on the nitrogen’s partial pressure at 3.8 Ga are insensitive to the specifics of the adopted \ce{CO2} evolutionary history in the wide range that has been explored. Therefore, an initially large \ce{N2} reservoir preferred by the isotopic constraints is fully compatible with an initially large \ce{CO2} reservoir.

One might ask if Mars started with nearly equal abundances of \ce{N2} and \ce{CO2} in the atmosphere, why does it currently have a \ce{CO2}-dominated atmosphere? The atmospheric composition is deflected towards \ce{CO2} dominance because \ce{N2} is much more prone to sputtering loss than \ce{CO2}. With equal mixing ratios in the bulk atmosphere, the sputtering loss rate of \ce{N2} is $\sim32$ times higher than that of \ce{CO2}, with a factor of $\sim8$ from diffusive enrichment and a factor of $\sim4$ from the difference in the sputtering yield (see Eq. \ref{eq:sput}). Thus, sputtering can quickly reduce the \ce{N2} abundance in the Noachian and Hesperian periods to establish the \ce{CO2}-dominated atmosphere. This preferential lost of \ce{N2} versus \ce{CO2} may well be a common feature of nonthermal escape from unmagnetized planets and might affect the evolutionary outcomes of terrestrial exoplanets.

Another question is how to form the nitrogen-rich atmosphere on Mars in the first place. The ratio between \ce{CO2} and \ce{N2} was probably $>10$ in volcanic outgassing\cite{mckay1989early,marty2003nitrogen,hirschmann2008ventilation}. Thus, we speculate that the atmosphere at 3.8 Ga primarily came from the late veneer of primitive bodies such as comets that have low C/N ratios\cite{bergin2015tracing}, after the atmosphere built by earlier volcanic outgassing had been removed by hydrodynamic escape\cite{tian2009thermal}. If the cometary origin was the case, the initial $\delta^{15}$N might be higher than what we have assumed ($\sim800\permil$\cite{marty2012origins}). Nonetheless, the evolution models and isotope-based constraints are not sensitive to this variation of the initial $\delta^{15}$N.

Measurements of future in-situ or sample return exploration can further delineate the steady-state versus dynamical nitrogen evolution. Fig.~\ref{fig:sample} shows that the $\delta^{15}$N value of the dynamical solutions has a broad peak between 3.5 and 1 Ga (i.e., the Hesperian and early Amazonian periods), and that of the steady-state solutions does not. The magnitude of the peaks with a non-collapsed atmosphere is $\sim5000-20000\permil$, and that with a collapsed atmosphere can be even larger. If some of this signal is transferred to the nitrates formed in the early Amazonian, isotopic analyses of nitrate samples could provide additional evidence for the dynamical evolution of nitrogen as well as the atmospheric collapse in the history of Mars.

Under the new interpretation of nitrogen's isotopic composition presented here, a few hundred mbar \ce{N2} may have existed in the Martian atmosphere at 3.8 Ga. Previous one-dimensional (1D) radiative-convective climate models have determined that this amount of additional nitrogen could cause the mean surface temperature to be $\sim10$ K warmer than a pure, 1-bar \ce{CO2} atmosphere via pressure broadening of \ce{CO2} absorption bands\cite{von2013n2}. The additional few hundred mbar \ce{N2}, coupled with substantial early carbonate deposition, may thus be particularly meaningful for sustaining surface melting at low topographic regions\cite{forget20133d}. The climate conditions on early Mars have been a conundrum\cite{ramirez2014warming,kite2017methane,wordsworth2017transient}, and our findings may provide a new path forward for explaining the geologic records\cite{grotzinger2015deposition} that suggest persisting liquid-water conditions on ancient Mars's surface.

\begin{figure}[!htbp]
\centering
\includegraphics[width=0.5\textwidth]{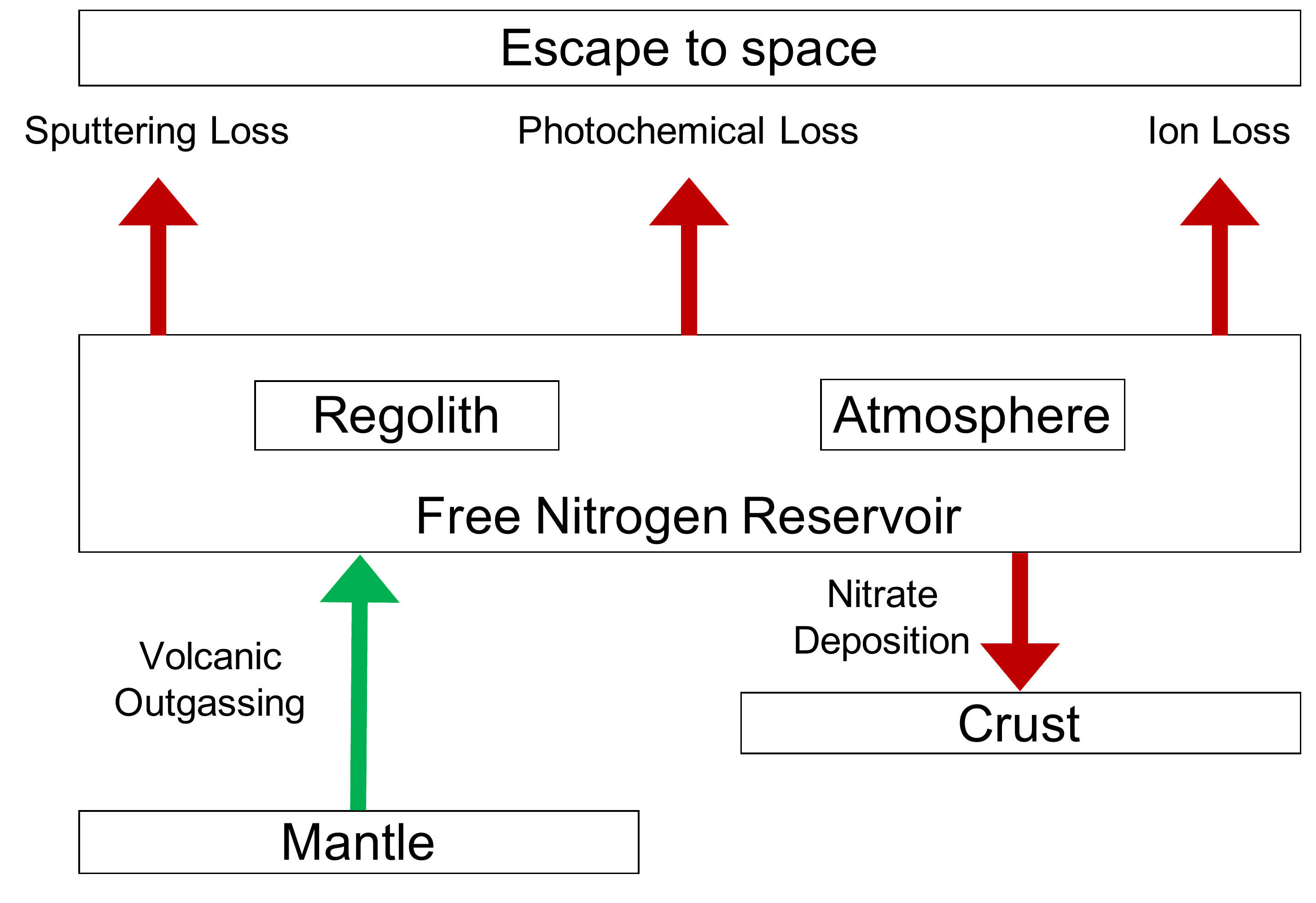}
\caption{A model for the long-term changes of the free nitrogen reservoir on Mars. The free nitrogen reservoir is comprised of \ce{N2} adsorbed in the regolith and \ce{N2} in the atmosphere, and it changes over time due to sputtering loss, photochemical loss, ion loss, volcanic outgassing, and nitrate deposition. The regolith and the atmosphere are assumed to exchange isotopes over geologic timescales driven by the temperature variations due to orbital obliquity changes\cite{laskar2004long}. We do not include impact additions or removal because the major impacts should have occurred before the modeled period\cite{fassett2011sequence,robbins2013large} (from 3.8 Ga to present). Nor do we include the impact decomposition of near-surface nitrates\cite{manning2008nitrogen} explicitly, as its rate is less than the present-day outgassing and photochemical escape rates by several orders of magnitude (Supplementary Information H).}
\label{fig:boxmod}
\end{figure}

\begin{figure}[!htbp]
\centering
\includegraphics[width=0.5\textwidth]{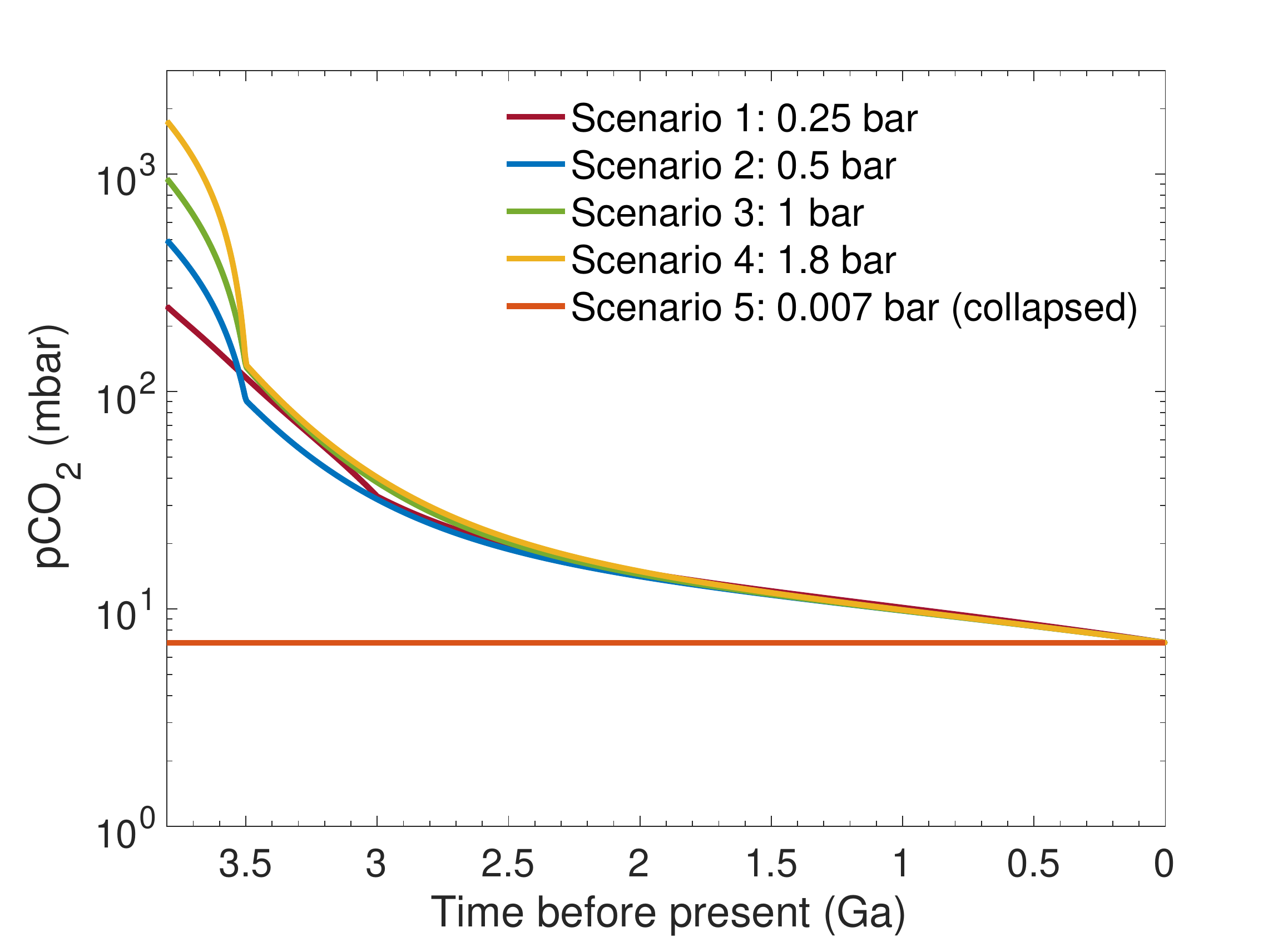}
\caption{Representative background \ce{CO2} evolutionary scenarios adopted in this study. These scenarios are selected from the evolution tracks of \ce{CO2} derived in Hu et al. (2015)\cite{Hu2015} and they are consistent with the present-day pressure and carbon isotopic composition. Scenarios 1 -- 3 assume that the photochemical loss rate depends on the Sun's Lyman continuum flux to the power of 2. Scenario 1 assumes that carbonate deposition of 40 mbar occurred throughout the Noachian and Hesperian (i.e., till 3.0 Ga) in shallow subsurface aquifers. This represents the lower bound of the initial \ce{CO2} partial pressure. Scenarios 2 and 3 assume that carbonate deposition of 290 and 600 mbar occurred in the Noachian (i.e., till 3.5 Ga) in open-water systems. Scenario 3 has an initial \ce{CO2} partial pressure of 1 bar and is the default scenario adopted in this study. Scenario 4 assumes a power-law index of 3, and that carbonate deposition of 1400 mbar occurred in the Noachian (i.e., till 3.5 Ga) in open-water systems. This represents the upper bound of the initial \ce{CO2} partial pressure. Scenario 5 is an endmember scenario where the \ce{CO2} atmosphere is assumed to be collapsed at all times and the pressure constant at 7 mbar.}
\label{fig:CO2}
\end{figure}

\begin{methods}

\subsection{Overview.}

The model starts at 3.8 Ga before the present and simulates the loss and addition of \ce{N2} in a combined reservoir of the atmosphere and the regolith (by adsorption) until the present day. It includes five addition and loss mechanisms shown in Fig.~\ref{fig:boxmod}. When tracing the evolution, the model uses a variable time step adapted to allow no more than 0.05\% change in the \ce{N2} reservoir per step. This ensures that the temporal resolution is high enough to capture important details of the atmosphere's evolution. 

An evolutionary solution must reach the present-day size and isotopic composition of the \ce{N2} reservoir. The size of the atmospheric reservoir is 0.12 mbar and that of the regolith reservoir is up to 0.01 mbar\cite{zent1994fractionation}. The upper limit corresponds to adsorption in up to 10 m global equivalent of the Martian regolith having $\sim20$ m$^2$ g$^{-1}$ surface area\cite{ballou1978chemical}. The isotopic composition of the present-day nitrogen reservoir is $\delta^{15}$N $=572\pm82\permil$\cite{Wong2013}. We do not include potential fractionation between the atmosphere and the regolith because the regolith reservoir size is minor compared with the atmosphere.

The initial size of the nitrogen reservoir is a free parameter of the model. The initial $\delta^{15}$N value of the reservoir, however, is uncertain. For the nominal cases, we assume the initial reservoir to have the isotopic composition of the mantle component of the nitrogen measured in the Martian meteorite ALH 84001\cite{mathew2001early} ($\delta^{15}$N$=-30\permil$). Nonetheless, escape and giant impacts that occurred before the modeled period may have modified the $\delta^{15}$N value from the mantle value, and previous estimates suggest that the $\delta^{15}$N value can vary rapidly between the mantle value and up to $500\permil$ more than 3.8 Ga before the present\cite{Kurokawa2018}. We thus explore the effect of an elevated initial $\delta^{15}$N value (e.g., $300\permil$, Supplementary Information F).

To calculate the mixing ratio of \ce{N2} (needed for calculating the escape rates), the model is coupled with an evolution model of \ce{CO2}\cite{Hu2015} for the same modeled period. The model of \ce{CO2} includes photochemical loss, sputtering loss, volcanic outgassing, and carbonate deposition\cite{Hu2015}, but not the feedback of the \ce{N2} mixing ratio on the escape rates of carbon. We present fully coupled carbon and nitrogen evolution models in Supplementary Information G. Otherwise, we explore nitrogen's evolutionary history by adopting the \ce{CO2} evolutionary scenarios with an initial partial pressure of \ce{CO2} ranging from 0.25 to 1.8 bar (Scenarios 1 -- 4 in Fig.~\ref{fig:CO2}). These scenarios are selected to represent a wide range of possible \ce{CO2} evolutionary tracks: in essence, carbon's isotopic composition constrains the relative strengths between photochemical loss and carbonate deposition, while the amount of carbonate deposition -- and thus the initial pressure -- can typically vary by a few hundred mbars due to the range in the possible timing and fractionation factor of carbonate deposition, as well as the uncertainties in the carbon isotope's measurement, outgassing models, and sputtering rates. The representative scenarios adopted here approximately encompass the lower and upper bounds of the \ce{CO2} partial pressure at 3.8 Ga allowed by the carbon isotopic composition\cite{Hu2015}.

We also include the endmember scenario of a collapsed atmosphere (Scenario 5 in Fig.~\ref{fig:CO2}). It has been suggested that the atmosphere of Mars can collapse to form large polar \ce{CO2} ice caps during low-obliquity periods and the remaining atmosphere would be in vapor pressure balance with the ice caps\cite{nakamura2003climate,forget20133d,soto2015martian}. The collapse of \ce{CO2} may matter for the evolution of nitrogen because it greatly increases the mixing ratio of \ce{N2} in the atmosphere. For example, the latest nitrogen evolution model\cite{Kurokawa2018} assumed that atmospheres with \ce{CO2} partial pressure $<500$ mbar collapse. Here we explore the effect of the collapse with an endmember scenario where the atmosphere is considered always `collapsed,' i.e., maintaining a \ce{CO2} partial pressure of 7 mbar.

\subsection{Photochemical Escape.} \label{sec:photochem}

The photochemical escape rate of nitrogen ($F_{\rm pr}$) is modeled as
\begin{equation}
    F_{\rm pr} = F_{\rm 0,pr} \bigg(\frac{F_{\rm EUV}}{F_{0, \rm EUV}}\bigg)^a \frac{X_{\ce{N2}}}{X_{0, \ce{N2}}} f_{\rm pr}, \label{eq:photochem}
\end{equation}
where the quantities with the subscript $_0$ are at the current epoch, $F_{\rm EUV}$ is the solar EUV flux for which we adopt as $F_{\rm EUV}\sim t^{-1.23}$ where $t$ is the age\cite{ribas2005evolution,claire2012evolution,tu2015extreme}, $X_{\ce{N2}}$ is the mixing ratio of \ce{N2} in the bulk atmosphere, $a$ is a power-law index, and $f_{\rm pr}$ is a multiplication factor. The photochemical escape rate is proportional to the mixing ratio of \ce{N2} in the atmosphere\cite{Fox1983}.
The power-law index and the multiplication factor provide sufficient freedom for the model to capture the uncertainties in the photochemical escape rate and how it has changed with the solar input.
This parameterization also consolidates the uncertainty in the age dependency of the solar EUV flux into $a$ (Supplementary Information F).

We break down the photochemical escape into major escape mechanisms because they have different isotopic fractionation factors, as
\begin{equation}
    F_{\rm pr} = F_{\rm photo} + F_{\rm recom} + F_{\rm chem},
\end{equation}
where $F_{\rm photo}$, $F_{\rm recom}$, and $F_{\rm chem}$ are the rates of escape produced by photodissociation and photoionization, dissociative recombination, and other chemical reactions, respectively. The current-epoch rates of these processes are baselined by the upper-atmosphere models of Fox (1993)\cite{Fox1993}, except for the photodissociation escape rate, which we reevaluated in Supplementary Information B. We assume that the scaling in Eq. (\ref{eq:photochem}) applies equally to these processes. Technically speaking, the photodissociation escape is driven by higher energy parts of the solar EUV flux (e.g., the Lyman continuum) and should have a different scaling. Applying the same scaling however is not a problem because we find that the photodissociation escape rate turns out to be minimal.

The isotopic fractionation factor ($\alpha$) of each of these processes is the product of the fractionation factor due to diffusive separation from the homopause to the exobase ($\alpha_{\rm diffu}$), and the fractionation factor when the escaping particles are launched by these processes near the exobase ($\alpha_{\rm photo/recom/chem}$), as
\begin{equation}
    \alpha = \alpha_{\rm diffu} \alpha_{\rm photo/recom/chem}.
\end{equation}
The diffusive separation is modeled as
\begin{equation}
    \alpha_{\rm diffu} = \exp\bigg(\frac{-g\Delta m\Delta z}{kT}\bigg), \label{eq:diffu}
\end{equation}
where $g$ is Mars's surface gravity, $\Delta m$ is the mass difference of the atoms or isotopologues in question, $\Delta z$ is the distance from the homopause to the exobase, $k$ is the Boltzmann constant, and $T$ is the mean temperature of the thermosphere. We include the effect of diffusive separation from the homopause all the way to the exobase, even though the peak of the photodissociation rate occurs well below the exobase\cite{bakalian2006production}, because only the energetic particles sourced at altitudes close to the exobase have a high probability to escape\cite{bakalian2006monte,cui2019photochemical}. It is evident from Eq. (\ref{eq:diffu}) that $\alpha_{\rm diffu}$ is controlled by the quantity $\Delta z/T$, which is constrained by MAVEN measurements of the argon isotopes (Supplementary Information D).

We determine $\alpha_{\rm photo}$ in Supplementary Information B, adopt previous calculations\cite{FoxHac1997} for $\alpha_{\rm recom}(=0.58)$, and assume $\alpha_{\rm chem}=1$. The main chemical reaction that produces escaping nitrogen is \ce{N2+ + O -> NO+ + N}. This reaction produces 10.9 eV as the excess energy, and this energy is partitioned into the products as the kinetic energy\cite{Fox1983}. The kinetic energy partitioned to \ce{N} is much higher than the escape threshold energy ($\sim1.73$ eV). This process therefore does not strongly fractionate nitrogen. Similarly, because dissociative ionization and electron impact dissociation produce energies more than 10 eV, these processes do not strongly fractionate nitrogen in the escape.

\subsection{Sputtering.}

Sputtering by oxygen ions picked up and propelled by the magnetic field of the solar wind can also cause escape. The loss due to sputtering is expected to be stronger at earlier epochs, when the solar EUV flux was higher. This process occurs when Mars does not have a strong magnetic field during the modeled period\cite{lillis2008rapid}, and the atmosphere is exposed to the solar wind. To model the sputtering loss rate of \ce{N2} ($F_{\rm sp}$), we scale the sputtering rate of \ce{CO2} ($F_{\rm sp}(\ce{CO_2})$) by the yields and mixing ratios\cite{Kurokawa2018,Jakosky1994,Slipski2016} of \ce{N2} and \ce{CO2} as
\begin{equation}
    F_{\rm sp}= F_{\rm sp}(\ce{CO_2}) \frac{Y_{\rm N_2}}{Y_{\rm CO_2}} \frac{X_{\rm N_2}}{X_{\rm CO_2}} \alpha_{\rm diffu, \ce{N2}/\ce{CO2}} \frac{1}{\alpha_{\rm dil}}f_{\rm sp},
    \label{eq:sput}
\end{equation} 
where $Y$ is the yield, $\alpha_{\rm diffu, \ce{N2}/\ce{CO2}}$ is the diffusion separation between \ce{N2} and \ce{CO2} calculated by Eq. (\ref{eq:diffu}) with $\Delta m$ equal to the mass difference between \ce{N2} and \ce{CO2}, $f_{\rm sp}$ is a multiplication factor to account for uncertainties in the flux, and $\alpha_{\rm dil}$ is a dilution factor that accounts for the dilution of \ce{N2} by other species at the exobase and is defined as
\begin{equation}
    \alpha_{\rm dil} \equiv 1 + \sum_i \frac{X_i}{X_{\rm CO_2}} \alpha_{\rm diffu, \textit{i}/CO_2},
\end{equation}
where $\alpha_{\rm diffu, \textit{i}/CO_2}$ is the diffusion separation between the species $i$ and \ce{CO2} defined similarly to Eq. (\ref{eq:diffu}). In addition to \ce{N2}, we include in this sum the minor species in Mars' atmosphere: \ce{Ar}, \ce{O}, and \ce{CO}, which currently have the abundances of 1.6\%, 0.13\%, and 0.08\% by volume respectively.

Let us consider the dependency of the sputtering rate on the size of the nitrogen reservoir. Expanding $\alpha_{\rm dil}$ in Eq. (\ref{eq:sput}), and only including the terms relevant to $X_{\ce{N2}}$, we have
\begin{equation}
    F_{\rm sp}\sim F_{\rm sp}(\ce{CO_2}) \frac{Y_{\rm N_2}}{Y_{\rm CO_2}}f_{\rm sp}
    \frac{\frac{X_{\rm N_2}}{X_{\rm CO_2}} \alpha_{\rm diffu, \ce{N2}/\ce{CO2}}}{1+\frac{X_{\rm N_2}}{X_{\rm CO_2}} \alpha_{\rm diffu, \ce{N2}/\ce{CO2}}} .
    \label{eq:sputdep}
\end{equation}
Therefore when $\frac{X_{\rm N_2}}{X_{\rm CO_2}} \alpha_{\rm diffu, \ce{N2}/\ce{CO2}}\ll1$ the scaled sputtering rate is proportional to the size of the nitrogen reservoir, and when $\frac{X_{\rm N_2}}{X_{\rm CO_2}} \alpha_{\rm diffu, \ce{N2}/\ce{CO2}}\gg1$ the sputtering rate no longer depends on the size of the nitrogen reservoir. Because $\alpha_{\rm diffu, \ce{N2}/\ce{CO2}}\sim8.5$ for a typical $\Delta z/T$ of 0.3 km/K, the sputtering rate becomes decoupled from the size of the nitrogen reservoir when it approaches the size of the carbon reservoir. This decoupling gives rise to slow convergence to the steady states and eventually the dynamical solutions.

We adopt the 3D Monte Carlo simulations\cite{leblanc2002} as the baseline values of the sputtering rate of \ce{CO2}, fitted to the following functional form
\begin{equation} 
    F_{\rm sp}(\ce{CO2}) = \exp(-0.462\ln(F_{\rm EUV}/F_{\rm 0,EUV})^2 + 5.086\ln(F_{\rm EUV}/F_{\rm 0,EUV}) + 53.49)
\end{equation}
where $F_{\rm sp}$ in this formula is the sputtering escape flux in particles per s. The basis for this formula comes from the product of the previously calculated fluxes of the incident pickup ions\cite{luhmann1992evolutionary} and the yield of escaping carbon particles\cite{leblanc2002}. The pickup ion fluxes calculated by Luhmann et al. (1992)\cite{luhmann1992evolutionary} for the current epoch are remarkably consistent with more sophisticated 3D models within a factor of two\cite{fang2013importance,wang2015statistical}, and also with the globally averaged precipitating ion flux and its energy spectrum measured by MAVEN\cite{leblanc2015mars}. For the sputtering yield of \ce{N2} we adopt the value calculated by the Monte Carlo simulations in Jakosky et al. (1994)\cite{Jakosky1994}, as it is the latest value. 

Lastly, the sputtering loss is energetic enough so that it does not preferentially select the escape of the N isotopes. The fractionation factor in the sputtering loss is the diffusive fractionation from the homopause to the exobase, i.e., $\alpha_{\rm diffu}$ given by Eq. (\ref{eq:diffu}). While the pickup ion's energy is principally deposited at altitudes well below the exobase, the recoil particles are quickly thermalized if they are produced at that altitude\cite{leblanc2001sputtering}. Only the particles produced near the exobase can escape. We thus use the exobase as the proxy for the source altitude of the sputtering loss. 

\subsection{Ion Loss.}

Ion loss is the only non-thermal escape process that has been directly measured at Mars\cite{barabash2007martian,jakosky2015maven,curry2018maven}, and the measured rates of carbon (as \ce{CO2+}) and oxygen (as \ce{O2+} and \ce{O+}) ion escape are generally consistent with magnetohydrodynamics model predictions\cite{ma2007ion}. Because the escape rate depends on age, we adopt a parametric model\cite{manning2011parametric} that fits the ion escape rates of the MAVEN-validated MHD model\cite{ma2007ion} to a power law of the age. The ion loss rate, $F_{\rm ion}$, is thus modeled as
\begin{equation}
    F_{\rm ion} =  \frac{X_{\ce{N2}}}{X_{0, \ce{N2}}} \frac{X_{0, \ce{N2+}}}{X_{0, \ce{CO2+}}} F_{\rm 0,ion}\big(\ce{CO2+}\big) \bigg( \frac{t}{4500} \bigg)^{-3.51},\label{eq:ion}
\end{equation}
where $X_{0, \ce{N2+}}$ and $X_{0, \ce{CO2+}}$ are the present-day mixing ratios measured by MAVEN at the altitude of 160 km\cite{bougher2015early}, $F_{\rm 0,ion}\big(\ce{CO2+}\big)$ is the present-day ion escape rate of \ce{CO2+} from the parametric model\cite{manning2011parametric}, and the last term accounts for the evolution of the solar EUV flux with the power-law index from the parametric model\cite{manning2011parametric}. Similar to sputtering, the fractionation factor of ion loss is $\alpha_{\rm diffu}$.

MAVEN observations indicated an approximately an-order-of-magnitude increase in the ion fluxes during a large interplanetary coronal mass ejection (ICME) event, while the observations took place at sparse locations\cite{jakosky2015maven}. The ensemble collections of MAVEN's ion observations during ICME events indicated that ICMEs result in a general decrease in the loss rate upstream, at low solar zenith angle (SZA), and an increase downstream, at higher SZA\cite{curry2018maven}. Integrating over SZA, these observations produced a net factor-of-two enhancement in the loss rate, although most of the events in this sample are small events. The young Sun likely produced more flare events and thus had a larger impact on the atmospheric loss. We do not include a multiplication factor in Eq. (\ref{eq:ion}) because the total amount of \ce{N2} lost from 3.8 Ga to the present day via ion loss is less than sputtering by several orders of magnitude. Thus, a variation in the ion loss rate by an order of magnitude would not change our model. We instead explore the sputtering multiplier $f_{\rm sp}$ in a wide range that captures the uncertainty in the pickup ion flux affected by solar activities.

\subsection{Nitrate Formation.}

See Supplementary Information C.

\subsection{Volcanic Outgassing.}

We estimate the volcanic outgassing flux ($F_{\rm og}$) from the history of crustal formation and the \ce{N2} content in the source magma. We model this as
\begin{equation}
    F_{\rm og} =  V\rho_{\rm cr} x_{\rm N_2} f_{\rm og},
\end{equation}
where $V$ is the crustal production rate, $\rho_{\rm cr}$ is the density of the crust, $x_{\rm N_2}$ is the concentration of \ce{N2} in the source magma, and $f_{\rm og}$ is a multiplication factor that accounts for the uncertainty in the crustal formation rate (including the extrusive-to-intrusive ratio) and the uncertainty in the outgassing efficiency.

For the crustal production rate, we combine the thermal evolution model estimate (i.e., the `global melt' scenario in Grott et al. 2011\cite{grott2011}) and the photogeological analysis of volcanic provinces on the planet's surface\cite{greeleyschneid1991} (Fig.~\ref{fig:crustal}). The outgassing factor in our evolution model $f_{\rm og}$ is directly comparable to the outgassing factor in the argon evolution model (i.e., the parameter $v_f$ in Slipski et al. 2016\cite{Slipski2016}). The crustal production rate in our model is highest at 3.8 Ga, and remains quite high throughout the Hesperian period. The total volume of volcanic emplacement between 3.8 and 3.0 Ga in our baseline model is approximately $4\times10^8$ km$^3$, compared to $3\times10^8$ km$^3$ as the estimated volume of the Tharsis Rise\cite{phillips2001ancient}. Our volcanic outgassing model is thus consistent -- within a factor of a few and captured by the multiplication parameter $f_{\rm og}$ -- with the recently revised picture that major volcanic emplacements such as the Tharsis Rise occurred during the late Noachian and the Hesperian\cite{tanaka2014digital,bouley2018revised}. Our volcanic model includes a non-negligible rate in the Amazonian, particularly within the last 500 Myr. This is also consistent with the geological evidence for active volcanism until very recently (0.1 Ga\cite{werner2009}).

Unlike carbon whose concentration in the magma is limited by its solubility\cite{hirschmann2008ventilation}, the concentration of \ce{N2} in the magma is typically not solubility-limited and thus reflects the formation history of the planet itself. In this work we adopt the `Silicate Earth' concentration\cite{marty2003nitrogen} ($x_{\rm N_2} = 1.9 \times 10^{-6}$) as the baseline value, and also recognize that Mars may have a different amount of nitrogen in the first place and may have a non-uniform distribution of nitrogen in the mantle. The uncertainty in $x_{\rm N_2}$ can be absorbed into the factor $f_{\rm og}$.

Volcanic outgassing introduces fresh nitrogen into the atmosphere and causes its isotopic composition to change with bulk mixing. The isotopic ratio, $\delta^{15}\ce{N}$, is modeled as
\begin{equation}
    \delta ^{15}\ce{N} = \frac{P_{\rm og}\delta^{15}\ce{N}_{\rm mantle} + P_{\rm atm}\delta ^{15}\ce{N}}{P_{\rm og}+P_{\rm atm}},
\end{equation}
where $P_{\rm og}$ is the partial pressure of \ce{N2} outgassed per time step, $P_{\rm atm}$ is the partial pressure of \ce{N2} in the bulk atmosphere, and $\delta^{15}\ce{N}_{\rm mantle}$ is the isotopic composition of nitrogen in the mantle. We assume $\delta^{15}\ce{N}_{\rm mantle}$ to be the value measured in the Martian meteorite ALH 84001\cite{mathew2001early}. Two isotopically distinct components of nitrogen have been found in ALH 84001\cite{mathew2001early}. The lighter one ($\delta^{15}\ce{N}=-30\permil$) may correspond to the nitrogen from the mantle, and the slightly evolved one $\delta^{15}\ce{N}=7\permil$ may correspond to the early Martian atmosphere.

\subsection{Markov-Chain Monte Carlo Simulations.} 

We employ the MCMC method to explore the vast parameter space of the evolution model. The likelihood function ($L$) defined for the MCMC analysis is
\begin{equation}
    \log L = 
    \left(\frac{P_{\rm 0,observed} - P_{\rm 0,model}}{\sigma_{\rm P}}\right)^2 + \left(\frac{\delta_{\rm 0,observed} - \delta_{\rm 0,model}}{\sigma_{\rm \delta}}\right)^2, \label{eq:likelihood}
\end{equation}
where $\delta$ represents $\delta ^{15}\ce{N}$ and $\sigma$ is the uncertainty for the current size and composition of the free nitrogen reservoir. Because the current size is $\sim0.12-0.13$ mbar including the regolith adsorption, we adopt the center value of $P_{\rm 0,observed} = 0.125$ mbar and the uncertainty of $\sigma_{\rm P} = 0.0017$ mbar. Essentially this gives the nonzero denominator required in the likelihood function, and the small value for $\sigma_{\rm P}$ means that any successful model must match with the present-day atmospheric pressure. For the isotopic composition we use the values measured by Curiosity's Sample Analysis at Mars instrument\cite{Wong2013}: $\delta_{\rm 0,observed}=572 \permil$ and $\sigma_{\rm \delta}=82 \permil$. 

The seven parameters listed in Table \ref{table:priors} are included as the free parameters in the MCMC simulations, and they have flat prior distributions in the ranges in which they are allowed to vary. For each MCMC simulation, two 2,000,000-element chains are produced starting from the parameters chosen independently and randomly within the allowed ranges. These two chains are then tested for convergence using the Gelman-Rubin method\cite{gelman1992inference}, and if converged, combined to derive the posterior distributions.

\end{methods}

\section*{Methods References}


\begin{addendum}
\item We thank Yuk L. Yung, Bethany Ehlmann, Bruce Jakosky, Hiroyuki Kurokawa, Curtis Manning, Robert Johnson, Matthias Grott, Fabrice Gaillard, Marek Slipski, and Cheuk-Yiu Ng for helpful discussions. This work was supported by NASA Habitable Worlds grant NNN13D466T, later changed to 80NM0018F0612. The research was carried out at the Jet Propulsion Laboratory, California Institute of Technology, under a contract with the National Aeronautics and Space Administration.
\item[Author Contribution] R.H. designed the study and the evolution model, interpreted the results, and wrote the manuscript. T.B.T. implemented the evolution model, carried out the simulations, and interpreted the results.
\item[Competing Interests] The authors declare no competing interests.
\item[Correspondence] Correspondence and requests for materials should be addressed to Renyu Hu~(email: renyu.hu@jpl.nasa.gov).
\item[Data Availability] The data needed to generate all figures in the main text (Figs. 1–3) and Extended Data Figs. 2 and 3 are publicly available at Zenodo (https://doi.org/10.5281/zenodo.5760095).
\item[Code Availability] The source code of the nitrogen evolution model and the associated configuration files used in this study are publicly available at Zenodo (https://doi.org/10.5281/zenodo.5760095).
\item[Supplementary Information] Supplementary Information is available for this paper.
\end{addendum}

\beginextended

\clearpage
\newpage
\section*{Extended Data}

\clearpage
\newpage
\section*{Extended Data}

\clearpage
\newpage
\section*{Extended Data}

\begin{figure}[!htbp]
\centering
\includegraphics[width=0.9\textwidth]{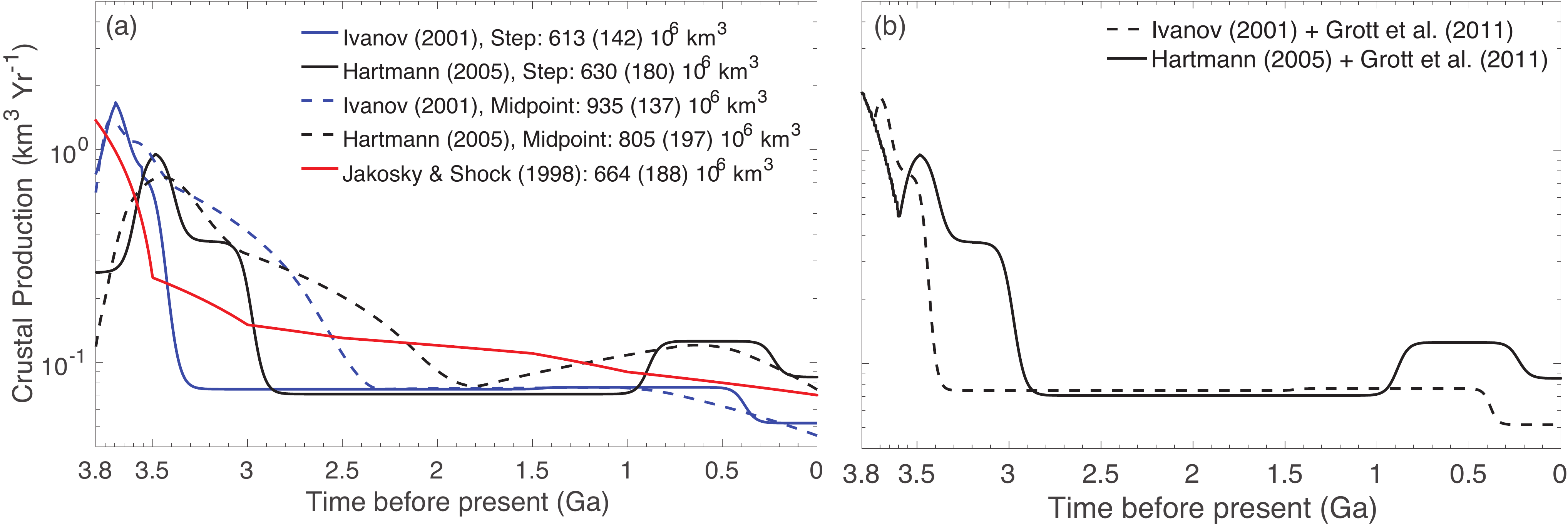}
\caption{Baseline crustal production rate adopted in this study. (a) Crustal production rates derived from the photogeologic analysis of volcanic provinces\cite{greeleyschneid1991} from 3.8 Ga. 
To convert the photogeological analysis (expressed as the total volcanic emplacement in each geologic period from middle Noachian to late Amazonian) to the crustal production rate, we compare the rates derived using the age boundaries from the crater density\cite{werner2011redefinition} and the chronology model of Ivanov (2001)\cite{ivanov2001mars} and Hartmann (2005)\cite{hartmann2005martian}, as well as interpolation methods using either step functions or mid-point averages. The labels show the total volcanic activity and the integrated volcanic activity in the last 2 billion years (in parentheses). The midpoint approach would introduce substantially more total volcanism, and so we use the step-function approach.
Also in comparison is a volcanic history derived from earlier photogeologic analyses and used in the recent argon isotope study\cite{Slipski2016}.
The step-function approach on the Hartmann (2015) chronology leads to a baseline model that is very similar to the one used in the argon isotope study\cite{Slipski2016} in terms of the total and the recent volcanic rates.
(b) Baseline crustal production rate adopted in this work, based on the global thermal evolution model\cite{grott2011} and the step-function interpolation of the photogeologic analysis of volcanic provinces\cite{greeleyschneid1991}, whichever is greater. We adopt the model using the Hartmann (2015) chronology as our baseline model, and consider the one using the Ivanov (2001) chronology as the variant.
The two models have appreciable difference in the last 2 billion years.}
\label{fig:crustal}
\end{figure}

\clearpage
\newpage
\section*{Extended Data}

\begin{figure}[!htbp]
\centering
\includegraphics[width=1.0\textwidth]{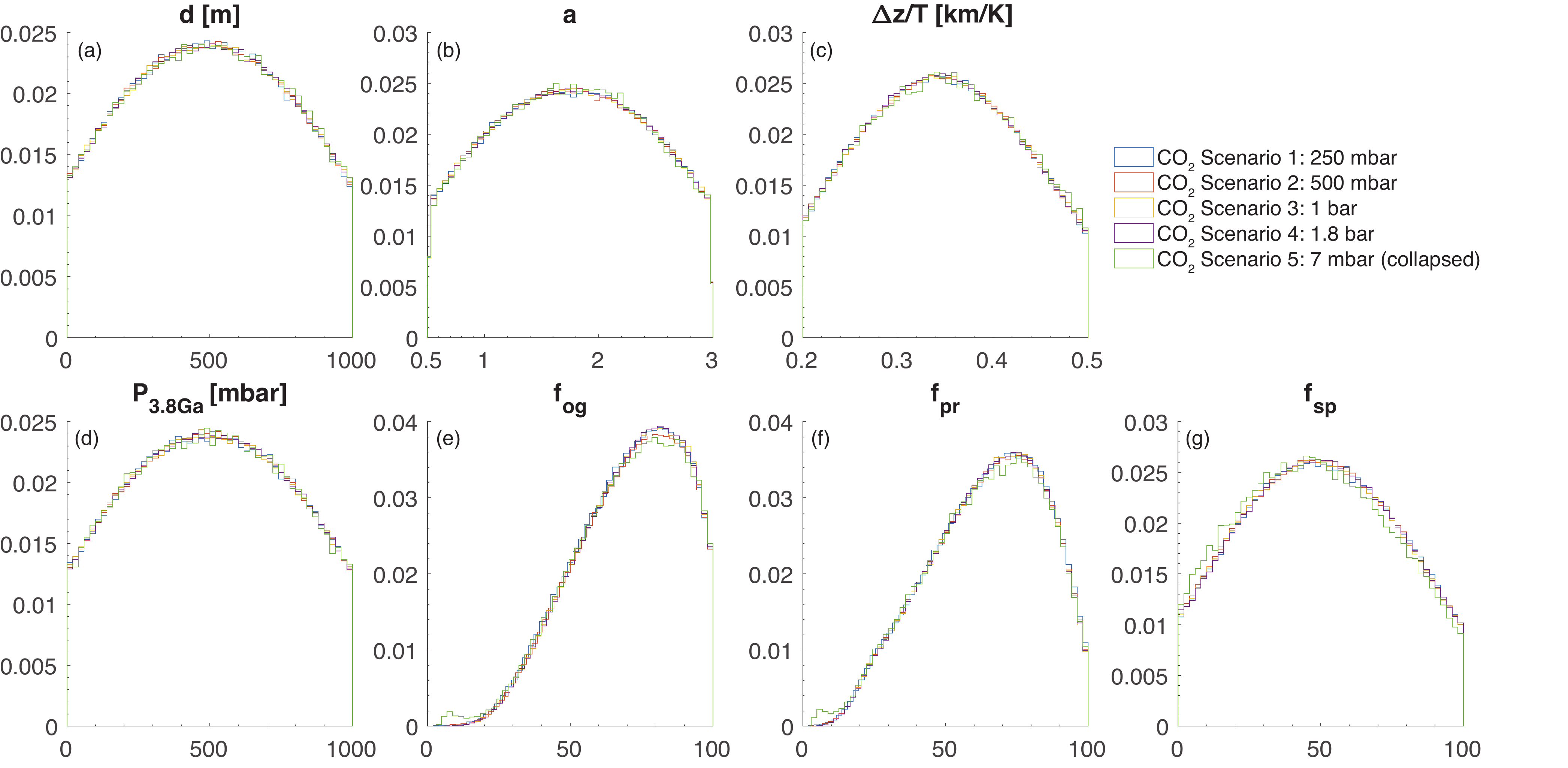}
\caption{Posterior distributions of parameters from unconstrained MCMC simulations with the parameters and their boundaries listed in Table~\ref{table:priors}. The MCMC simulations adopt the five representative \ce{CO2} evolution scenarios shown in Fig.~\ref{fig:CO2}.}
\label{fig:post_100}
\end{figure}

\clearpage
\newpage
\section*{Extended Data}

\begin{figure}[!htbp]
\centering
\includegraphics[width=1.0\textwidth]{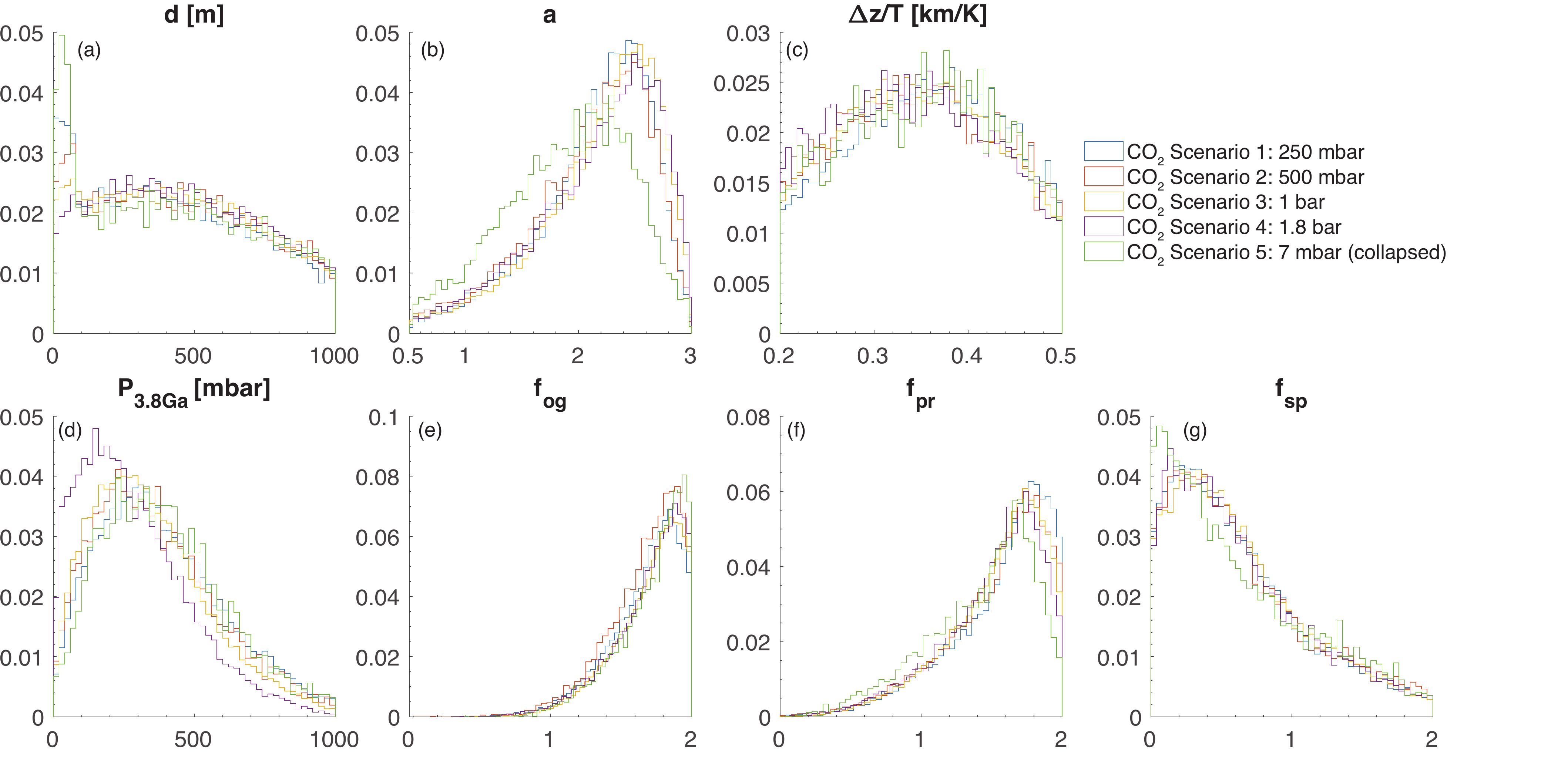}
\caption{Posterior distributions of parameters from constrained MCMC simulations with the parameters and their boundaries listed in Table~\ref{table:priors}. The MCMC simulations adopt the five representative \ce{CO2} evolution scenarios in Fig.~\ref{fig:CO2}.}
\label{fig:post_2}
\end{figure}

\clearpage
\newpage
\section*{Extended Data}

\begin{figure}[!htbp]
\centering
\includegraphics[width=1.0\textwidth]{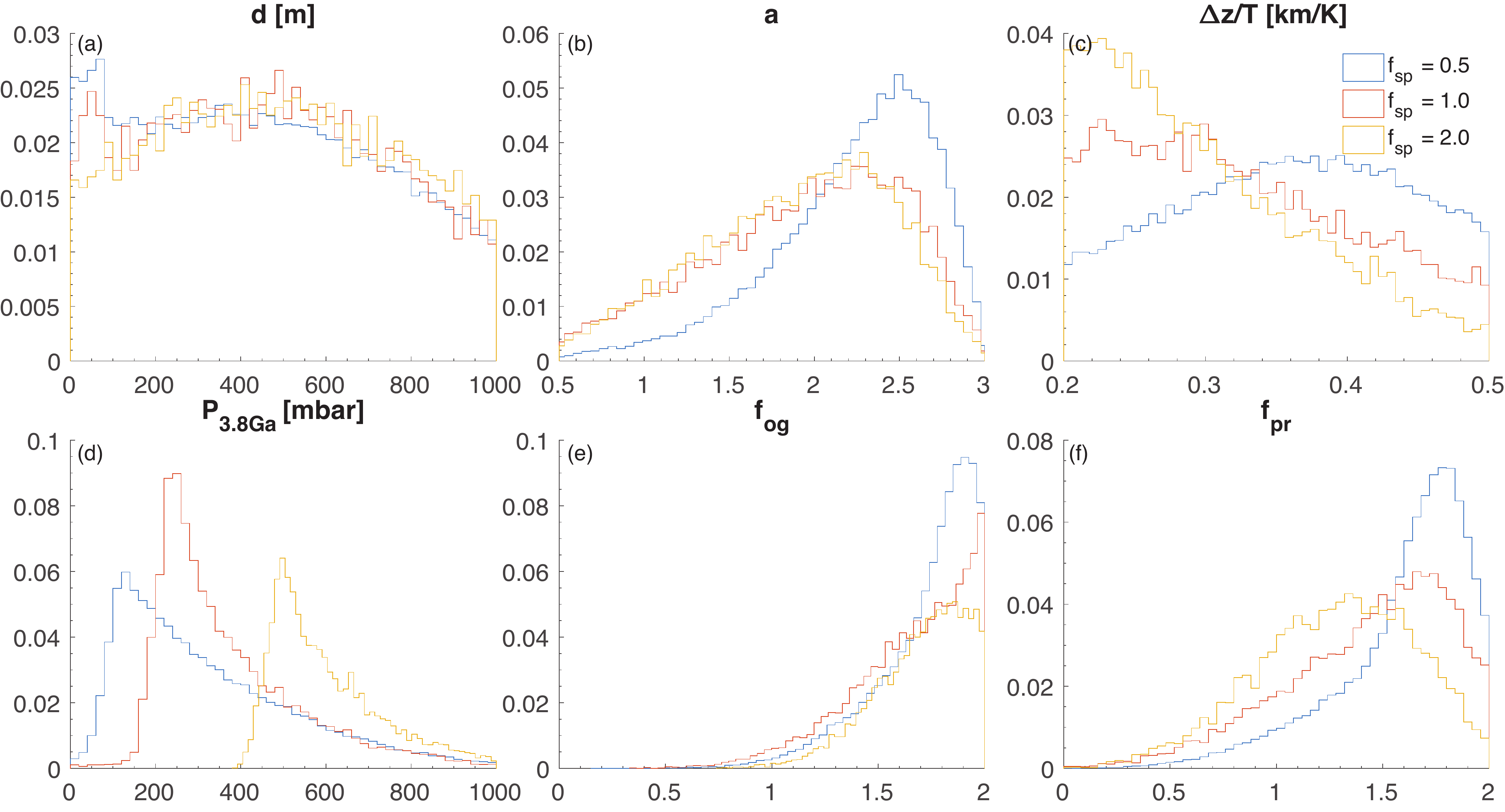}
\caption{Posterior distributions of parameters from constrained MCMC simulations that fix the sputtering multiplier to be $f_{\rm sp}=0.5$, 1, and 2. These simulations adopt the \ce{CO2} evolutionary scenario No. 3 with the initial partial pressure of 1.0 bar as shown in Fig.~\ref{fig:CO2}. The initial partial pressure of \ce{N2} is more tightly constrained when the sputtering multiplier is fixed.}
\label{fig:post_sp_multi}
\end{figure}

\beginsupplement
\nolinenumbers

\clearpage
\newpage
\setcounter{page}{1}

\section*{Supplementary Information for \\``A nitrogen-rich atmosphere on ancient Mars consistent with isotopic evolution models''}

Renyu Hu and Trent B. Thomas

\noindent Jet Propulsion Laboratory, California Institute of Technology, Pasadena, CA 91109

\noindent renyu.hu@jpl.nasa.gov

\section*{Supplementary Information A: Summary of Previous Evolutionary Models of Mars Nitrogen}

Table~\ref{table:nit_problems} summarizes the previously proposed solutions to explain Mars's nitrogen isotopic composition and their unresolved issues.

To explain the initial measurement of the nitrogen's isotopic composition by Viking\cite{mcelroy1976isotopic}, McElroy et al. (1976)\cite{McElroy1976} firstly modeled the photochemical escape of nitrogen and found that the escape caused much more fractionation than what was measured (Table~\ref{table:nit_problems}). The proposed solution for this over-enrichment problem is to have a large amount of \ce{N2} deposited as nitrates. However, the work did not include sputtering loss or the enhanced photochemical escape when the solar EUV flux was higher.

Fox \& Dalgarno (1983)\cite{Fox1983} and Fox (1993)\cite{Fox1993} substantially improved the estimates of the photochemical escape rate with realistic upper-atmosphere models and determined its dependency on the mixing ratio of \ce{N2} in the atmosphere. These papers showed that the photochemical escape would cause substantial over-enrichment, and proposed that an initial \ce{CO2} pressure of $1\sim2$ bar that diminished exponentially with a time constant in the range of 0.7 -- 1 billion years to the value today could reproduce the measured isotopic ratio of nitrogen. Still, these papers did not include sputtering loss or the enhanced photochemical escape when the solar EUV flux was higher.

Jakosky et al. (1994) presented the first comprehensive model of nitrogen evolution that includes sputtering loss and photochemical loss and their dependency on the evolving solar EUV flux. Their standard model reproduced the present-day nitrogen isotopic composition with a combination of the reduced sputtering rate (25\% of Luhmann et al. 1992\cite{luhmann1992}), an initial \ce{CO2} pressure of 0.75 bar, and a substantial and temporally spread outgassing profile that peaks at 3 Ga and extends well into 1 -- 2 Ga. A collapsed atmosphere scenario where the partial pressure of \ce{CO2} was kept at 7 mbar was considered, and that scenario required an even smaller sputtering rate. In these models, nitrogen is in a steady state, where sputtering balances outgassing, until $\sim1.5$ Ga and then deviates from the steady state due to the reduction of outgassing. This is, to our knowledge, the first time that the steady state in Mars nitrogen's evolution has been shown. Without the substantial outgassing, however, 1 -- 3 bar initial \ce{CO2} would be required to mitigate the over-enrichment problem. While the reduced sputtering rate may be plausible given the uncertainties in the sputtering models, the massive outgassing that extends well into the Amazonian appears to be at odds with the current geologic mapping\cite{werner2009}. The total amount of \ce{N2} outgassing required is 16 mbar, but more recent argon isotope measurements and models indicate that the total outgassing should be $<\sim1$ mbar\cite{Slipski2016} if Mars has the same N/Ar ratio as Earth. Without the outgassing, the required massive early \ce{CO2} atmosphere is not supported by the lack of widespread carbonate deposits on the Martian surface or the atmospheric carbon's isotopic composition\cite{edwards2015carbon,Hu2015}.

Zent et al. (1994)\cite{Zent1994} presented the experimental data for the \ce{N2} adsorption capacities of Martian regolith and assessed whether regolith adsorption of \ce{N2} can mitigate the over-enrichment problem as an alternative to the thick \ce{CO2} atmosphere and substantial volcanic outgassing. They determined that 2.4 mbar \ce{N2} adsorbed by the regolith would be required, which is about two orders of magnitude greater than a plausible amount.

Fox \& Ha\'{c} (1997)\cite{FoxHac1997} determined the isotopic fractionation factor in the dissociative recombination of \ce{N2+}, one of the main photochemical loss mechanisms. Without including sputtering, they suggested that the present-day isotopic composition can be matched by a moderate-size early \ce{CO2} atmosphere that has a time constant of more than 1 billion years.

Finally, Kurokawa et al. (2018)\cite{Kurokawa2018} presented the most up-to-date and comprehensive analysis of the nitrogen's isotopic composition together with the evolution of \ce{CO2} and noble gases. Their models traced the evolution from 4.5 Ga (compared to 3.8 Ga in Jakosky et al. 1994\cite{Jakosky1994} and this work) and included impacts of asteroids and comets. Their models assumed the atmosphere collapse if the atmospheric pressure is $<0.5$ bar. The preferred nitrogen evolution scenario features the steady state between the outgasssing and the sputtering until $\sim0.5$ Ga, and after that, the outgassing stops and the photochemical escape raises $\delta^{15}$N to the final value. The cessation of outgassing at $\sim0.5$ Ga is inconsistent with the geological evidence for active volcanism until very recently (0.1 Ga\cite{werner2009}). We suspect that this assumption of Kurokawa et al. (2018) might have originated from taking the data in Fig.~2 of Craddock \& Greeley (2009)\cite{craddock2009minimum} as the face value. The outgassing values drawn at the midpoint of each geologic period in Fig.~2 of Craddock \& Greeley (2009) are the mean flux for the period and thus should be applied to the whole period.

\begin{sidewaystable}[!htbp]
\centering
\begin{tabular}{llp{0.8cm}p{8cm}p{9cm}}
\hline \hline
Paper  & $\delta^{15}$N  (\textperthousand) & EUV & Proposed Solutions & Unresolved Issues\\
\hline \hline
McElroy et al. 1976 \cite{McElroy1976} & 1440 & No & $\sim10$ mbar \ce{N_2} deposited as nitrates & Did not include enhanced escape at early epochs \\ 
\hline
Fox \& Dalgarno 1983 \cite{Fox1983} & 1510 & No & Initial $\sim1$ bar \ce{CO_2} lasting for $0.7\sim1$ Ga & Did not include enhanced escape at early epochs \\
\hline
Fox 1993 \cite{Fox1993} & 1530 & No & Initial $1\sim2$ bar \ce{CO_2} lasting for $0.7\sim0.8$ Ga & Did not include enhanced escape at early epochs \\
\hline
Jakosky et al. 1994 \cite{Jakosky1994} & N/A & Yes & Initial 0.75 bar \ce{CO2} and large outgassing after 3.0 Ga (total outgassing $\sim$ 16 mbar \ce{N2}), or initial $1\sim3$ bar \ce{CO2} without outgassing & Outgassing profile and amount inconsistent with geologic mapping and argon isotopes, and massive \ce{CO2} inconsistent with the lack of carbonates \\
\hline
Zent et al. 1994 \cite{Zent1994} & 3900 & Yes & 2.4 mbar \ce{N_2} adsorbed in regolith & Not achievable for reasonable thickness of regolith \\
\hline
Fox \& Ha\'{c} 1997 \cite{FoxHac1997} & N/A & Yes & Initial 0.26 bar \ce{CO_2} lasting for 1.1 Ga & Did not include sputtering \\
\hline
Kurokawa et al. 2018 \cite{Kurokawa2018} & N/A & Yes & Steady state until 0.5 Ga followed by a cessation of outgassing & Inconsistent with geologic evidence of volcanism active until 0.1 Ga \\
\hline
\end{tabular}
\caption{Previous studies of the evolution of Martian nitrogen. $\delta^{15}$N is the estimate of the present-day isotopic composition calculated by the paper, before applying any proposed solutions to mitigate the over-enrichment problem. Typically this value was obtained by including the modeled escape processes only. The EUV column indicates whether or not the paper took into account the impact of the evolving solar EUV flux on the escape rate.}
\label{table:nit_problems}
\end{sidewaystable}

\clearpage
\newpage

\section*{Supplementary Information B: Fractionation Factor in the Escape Driven by Photodissociation of \ce{N2}}

The photodissociation of \ce{N2} has been recognized as a mechanism to produce escaping nitrogen on Mars\cite{brinkmann1971}. It is also known that photodissociation of \ce{N2} does not produce two nitrogen atoms in the ground state N($^4$S), and previous works have assumed that it produces one N in the ground state, and the other N is the first excited state N($^2$D)\cite{Fox1993,bakalian2006production}. Here we determine the rate and the fractionation factor of the escape driven by the photodissociation of \ce{N2} on Mars using the ``Photochemical Isotope Effect'' (PIE) method\cite{Hu2015}. We also incorporate the latest experimental data\cite{song2016quantum} that suggest the production of higher excited states and a complex relationship between the quantum yields of the dissociation channels and the energy of the incident photon (Table~\ref{table:n2}). 

In short, the PIE method distributes the energy of the incident photon that is more than the threshold energy of each channel into the two nitrogen atoms as the kinetic energy. This calculation uses the solar spectrum, the cross-sections of \ce{N2} photodissociation, and the quantum yields as a function of wavelength as the input. We use the data compilation of Fennelly \& Torr (1992)\cite{fennelly1992photoionization} for the cross-sections of photoabsorption and photoionization. Two energy limits are noteworthy: (1) photons with energy higher than $\sim 15.5$ eV (or wavelength $<79.8$ nm) cause ionization in addition to dissociation; (2) photons with energy higher than $\sim18.8$ eV (or wavelength $<66.0$ nm) only cause ionization.

\begin{sidewaystable}[!htbp]
\centering
\begin{tabular}{lllll} 
 \hline\hline
 No. & Channel & Threshold Energy & Min. Photon Energy & Quantum yields for photons with energy $E$ \\ 
 \hline\hline
(1) & N($^4$S) + N($^4$S) & 9.76 eV & 13.22 eV & 0 \\
(2) & N($^4$S) + N($^2$D) & 12.14 eV & 15.60 eV & 1 for $12.14<E<13.33$ eV, 0.6 for $13.33<E<14.53$ eV, \\
&&&& 0.4 for $14.53<E<15.15$ eV, unknown for $E>15.15$ eV\\
(3) & N($^4$S) + N($^2$P) & 13.33 eV & 16.79 eV & 0.4 for $13.33<E<14.53$ eV, 0 for $14.53<E<15.15$ eV, \\
&&&& unknown for $E>15.15$ eV\\
(4) & N($^2$D) + N($^2$D) & 14.53 eV & 17.99 eV & 0.6 for $14.53<E<15.15$ eV, unknown for $E>15.15$ eV \\
(5) & N($^2$D) + N($^2$P) & 15.72 eV & 19.18 eV & unknown \\
 \hline
\end{tabular}
\caption{Channels and their quantum yields of the photodissociation of \ce{N2}. The minimum photon energy to drive escape is the threshold energy plus the energy required for two $^{14}$N to escape from Mars's exobase ($=1.73\times2=3.46$ eV). The quantum yields are from Song et al. (2016)\cite{song2016quantum}.}
\label{table:n2}
\end{sidewaystable}

\begin{figure}[!htbp]
\centering
\includegraphics[width=0.6\textwidth]{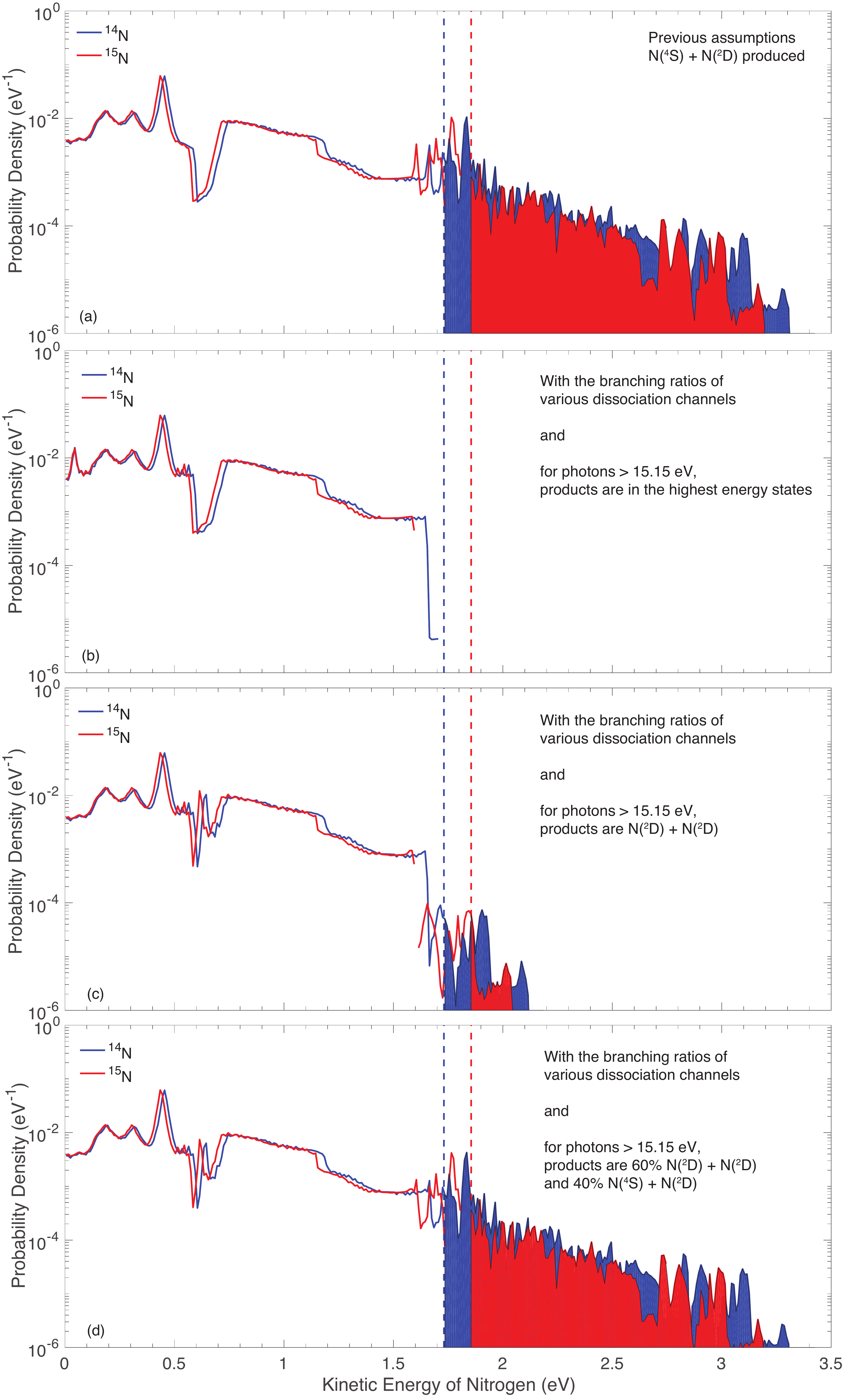}
\caption{Energy distribution of nitrogen atoms produced by the photodissociation of \ce{N2} in the upper atmosphere of Mars. The required energy for each isotope to escape is shown by dashed lines for comparison. The blue and red areas indicate the fraction of $^{14}$N and $^{15}$N that escapes, respectively. The panels show results from four assumptions on the quantum yields of the photodissociation channels, including the previous assumption\cite{Fox1993,bakalian2006production} and three scenarios incorporating the latest experimental data\cite{song2016quantum}.}
\label{fig:n2photo}
\end{figure}

Using the previous assumption\cite{Fox1993,bakalian2006production} that all the photons more energetic than 12.14 eV produce N($^4$S) and N($^2$D), we find that 6.03\% of the $^{14}$N and 1.75\% of the $^{15}$N produced in the photodissociation would escape (Fig. \ref{fig:n2photo}). For the first time we determine the fractionation factor to be $\alpha_{\rm photo}=0.29$.

With the possibility to produce the higher energy dissociation channels listed in Table \ref{table:n2}, the rate of escape is much reduced compared with the previous estimate\cite{Fox1993}. Because we do not have experimental data for the quantum yields from photons more energetic than 15.15 eV, we consider the following three scenarios to calculate the escape rate and the fractionation factor (Fig. \ref{fig:n2photo}). (1) First, the produced nitrogen atoms are always in the highest energy states allowed. In this case, it is evident from Table \ref{table:n2} that no produced nitrogen atoms would have enough energy to escape. (2) Second, the photons more energetic than 15.15 eV produce two N($^2$D)s, the highest-energy dissociation channel observed in the laboratory. In this case, only the photons with energy between 17.99 eV (see Table \ref{table:n2}) and 18.8 eV (i.e., the upper limit of the energy of the photon that causes dissociation) can drive escape. We find that only 0.06\% of the produced $^{14}$N have enough energy to escape (or 1\% of the result using previous assumption), and the fractionation factor would be $\alpha_{\rm photo}=0.15$. The process is highly selective between the isotopes because only the incident photons in the narrow range of energy can produce escape. The whole process nonetheless stops being important for evolution because the escape rate is too small. (3) Third, we extrapolate the measured quantum yields between 14.53 and 15.15 eV (Table \ref{table:n2}) to higher energies. In this case, we find that 2.5\% of the produced $^{14}$N have enough energy to escape (or 41\% of the result using previous assumption), and the fractionation factor would be $\alpha_{\rm photo}=0.29$. For the evolution model, we scale the previous estimate\cite{Fox1993} by the new yields for $F_{\rm photo}$ and use the newly determined fractionation factor for $\alpha_{\rm photo}$, and thus complete the treatment for photochemical escape.

To summarize, because more energetic photons produce nitrogen atoms in increasingly higher excited states, much of the energy from the solar irradiation does not drive escape. We find that $0\sim2.5$\% of the $^{14}$N produced by the photodissociation can escape, which is maximally 40\% of the previous estimate\cite{Fox1993}, depending on how the experimental data are extrapolated to higher-energy photons (Fig. \ref{fig:n2photo}). If the produced nitrogen atoms are always in the highest energy states allowed, including a never-observed channel of N($^2$D)+N($^2$P), none of the produced atoms would have enough energy to escape. If some nitrogen escapes, the PIE in this process is severe because the escaping particles are produced in a very narrow energy window, and the fractionation factor is $\alpha_{\rm photo}=0.15\sim0.29$.

\clearpage
\newpage

\section*{Supplementary Information C: Constraints of the Amount and Fractionation Factor of Nitrate Formation}

Evolved gas experiments with SAM on Curiosity have determined the concentration of nitrate in the Martian soil and rocks\cite{stern2015,sutter2017evolved}. Here we adopt the concentration measured in the Rocknest samples as representative of the Martian soil ($180\pm80$ ppm by weight in \ce{NO3}\cite{sutter2017evolved}). For a globally average regolith depth of 10 m, this corresponds to $0.03\pm0.01$ mbar of \ce{N2}. We evenly distribute this amount in the Amazonian period to estimate the removal rate of nitrogen from the free reservoir in this period. For the nitrate content in Noachian and Hesperian rocks, the SAM measurements have suggested highly variable nitrogen concentration. The weighted average of all rock samples measured by SAM is $88\pm13$ ppm by weight in \ce{NO3}, and the weighted average of only the Cumberland samples is $550\pm150$ ppm\cite{sutter2017evolved}. Also, the thickness of the crust affected by nitrate deposition is essentially unknown. Because the concentration and the depth are degenerate for our purpose, we use a default rock concentration of 300 ppm by weight in \ce{NO3}, and explore an equivalent depth ($d$) as a free parameter that can be as large as $\sim1000$ m. We evenly distribute the nitrate amount determined by the parameter $d$ in the modeled Noachian and Hesperian period (i.e., 3.8 -- 3.0 Ga) to estimate the removal rate of nitrogen from the free reservoir in this period. As a guiding estimate, the nitrate concentration in 500 m of crust corresponds to $\sim4$ mbar of \ce{N2}, larger than the size of the current reservoir by more than one order of magnitude. The crustal sequestration is thus potentially a significant loss for the nitrogen evolution in the atmosphere.

To estimate the fractionation factor for the nitrate formation we consider the formation processes. The nitrate formation must start from energetic processes that break \ce{N2} into \ce{NO}, and the relevant energetic processes include lightning, bolide impact, and high-temperature volcanism\cite{yung1979fixation,kasting1981limits,mckay1988high,mather2004volcanic,navarro2019abiotic}. Subsequent conversion from \ce{NO} to nitrates probably involved photochemical oxidation in the atmosphere\cite{smith2014formation} or conversion to \ce{HNO} and then disproportionation in the dissolved phase on or beneath the surface\cite{mancinelli1988evolution,summers2007nitrogen,hu2019stability}. The latter pathway requires alkaline solutions with pH $>11$\cite{hu2019stability} and may be consistent with the early Martian subsurface environments\cite{ehlmann2011subsurface}. We do not expect significant fractionation between the \ce{NO} produced by the energetic processes and the background \ce{N2}. Then, is there fractionation in the conversion from \ce{NO} to nitrate? Terrestrial measurements indicate that $\delta^{15}$N in \ce{HNO3} or particulate nitrates can be greater than that of the source atmospheric \ce{NOx} (i.e., \ce{NO} and \ce{NO2}) by up to $\sim10\permil$, and most of this fractionation probably comes from the isotope exchange reaction between \ce{NO} and \ce{NO2}\cite{freyer1991seasonal,freyer1993interaction,jarvis2008influence,chang2018nitrogen}. This empirical datum can be considered as a proxy for the fractionation in the photochemical oxidation pathway, and meanwhile, the fractionation in the \ce{HNO} pathway is essentially unknown. In this work, we consider no fractionation for nitrate formation in the standard scenarios and discuss the impact of a $10\permil$ fractionation (i.e., $\alpha_{\rm nitrate}=1.01$) as a variation. The $10\permil$ fractionation in nitrate deposition can only lead to $\sim1\permil$ change in the final $\delta^{15}$N (Fig.~\ref{fig:example}).

\clearpage
\newpage

\section*{Supplementary Information D: Constraints of $\Delta z/T$}

The quantity $\Delta z/T$ denotes the difference in altitude between the homopause and the exobase divided by the temperature of the atmosphere at this interval. This parameter is important to the diffusive fractionation as well as the sputtering loss rate. In the previous atmospheric evolution models\cite{Jakosky1994,Kurokawa2018}, this parameter was either adopted to be a single value or chosen as several values for different epochs corresponding to the varying solar EUV flux. Recently, MAVEN has measured this parameter on Mars using argon as a tracer\cite{jakosky2017}, and found that this parameter changes in the range of $0.2-0.5$ km K$^{-1}$ in a short period. This range also encompasses the entire range adopted in the previous evolution models. We therefore consider $\Delta z/T$ as a free but constant parameter that varies in $0.2-0.5$ km K$^{-1}$ in this work.

\clearpage
\newpage

\section*{Supplementary Information E: Correlated Parameters for the Steady-State Solutions}

\begin{figure}[!htbp]
\centering
\includegraphics[width=0.5\textwidth]{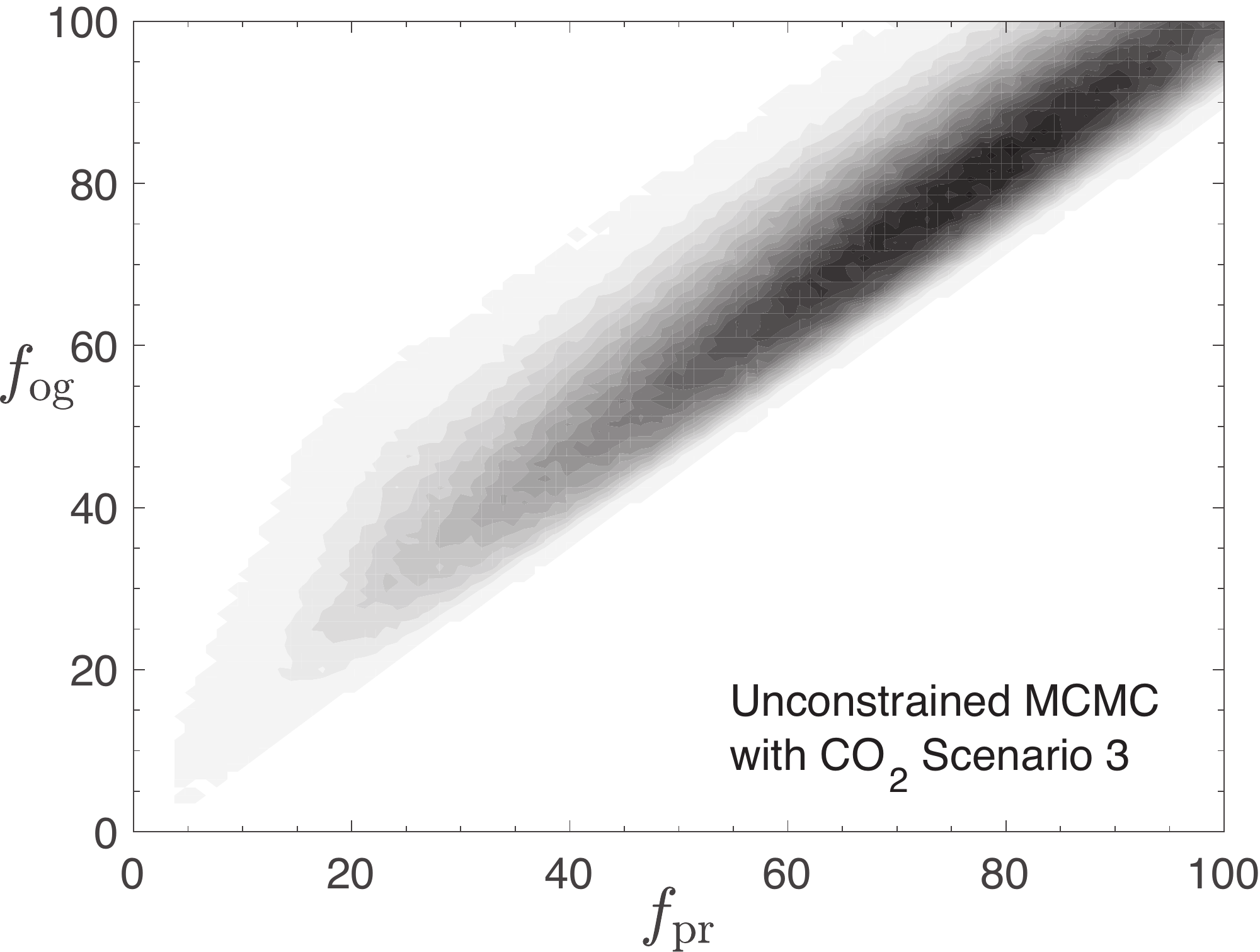}
\caption{Correlation between the multiplier of photochemical loss ($f_{\rm pr}$) and that of outgassing ($f_{\rm og}$), from the unconstrained MCMC simulation that adopts the \ce{CO2} evolutionary scenario No. 3 as shown in Fig.~\ref{fig:CO2}. The two multipliers are linearly correlated because the steady-state solutions require the photochemical loss rate to largely balance the outgassing rate at the present day.}
\label{fig:cor}
\end{figure}

The steady-state solutions can typically be found with large escape and outgassing rates that fit the correlation displayed in Fig.~\ref{fig:cor}. This is because the present-day $\delta^{15}$N value allows the current atmosphere to be in the steady state. The steady-state isotopic composition of atmospheric nitrogen is $1+\delta=(1+\delta_0)/\alpha$, where $\delta_0$ is the isotopic composition of outgassing and $\alpha$ is the end-to-end fractionation factor of escape\cite{McElroy1976}. Using the measured values of $\delta$ and $\delta_0$, $\alpha$ must be $0.62\pm0.03$ to allow the steady-state solutions. $\alpha$ of sputtering is from diffusive fractionation and is $0.87\pm0.04$ given MAVEN measurements\cite{jakosky2017}. $\alpha$ of photochemical loss includes additional fractionation due to PIE (Methods) and electron energy distributions\cite{FoxHac1997}, and is $0.64\pm0.03$. Therefore, the steady-state solutions require the current escape to be almost entirely made of photochemical loss. Meanwhile, the baseline outgassing, photochemical loss, and sputtering loss rates are $1.2\times$, $1.1\times$, and $0.3
\times10^{-4}$ mbar Myr$^{-1}$ at the present day; as a result, the following relationship should hold for mass balance: $1.2f_{\rm og}\sim1.1f_{\rm pr}+0.3f_{\rm sp}$. 
Putting the analyses together, the steady-state solutions would exist when $f_{\rm og}\sim f_{\rm pr} > f_{\rm sp}$. Thus, the solutions found by the unconstrained MCMCs (mostly steady-state solutions) often have large $f_{\rm og}$, $f_{\rm pr}$, and $f_{\rm sp}$.

\clearpage
\newpage

\section*{Supplementary Information F: Sensitivity of Posterior Distributions on Input Parameters}

\begin{figure}[!htbp]
\centering
\includegraphics[width=1.0\textwidth]{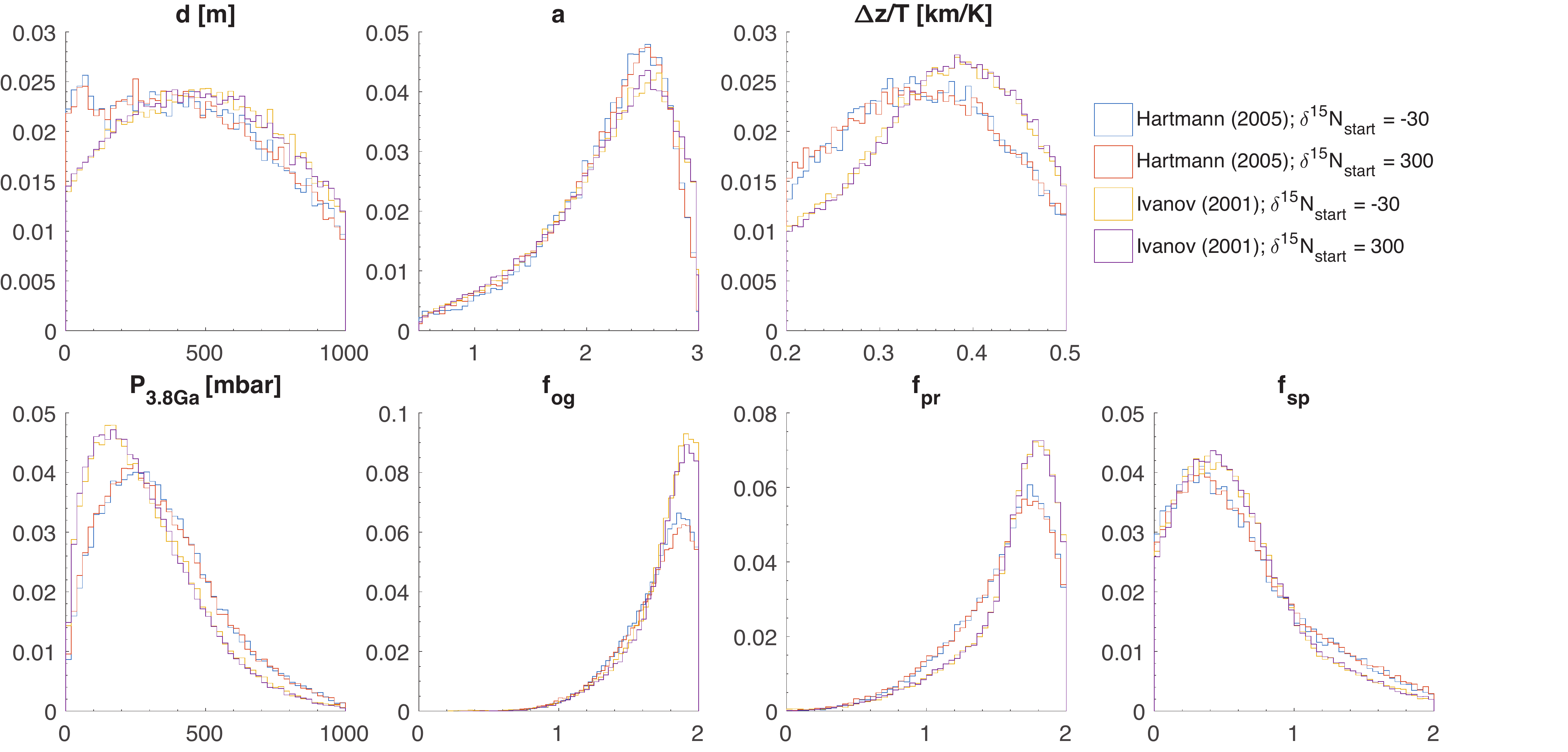}
\caption{Posterior distributions from constrained MCMC simulations that adopt two initial values for $\delta^{15}$N and two baseline outgassing models. The two baseline outgassing models are shown in Fig.~\ref{fig:crustal}, and their difference is driven by the crater chronology model of Ivanov (2001)\cite{ivanov2001mars} or Hartmann (2005)\cite{hartmann2005martian}. These simulations adopt the \ce{CO2} evolution scenario No. 3 with the initial partial pressure of 1.0 bar as shown in Fig.~\ref{fig:CO2}. The posterior distributions are not sensitive to the specific choices of the initial value of $\delta^{15}$N, and the difference in the baseline models for outgassing is largely accounted for by the photochemical loss parameters ($a$ and $f_{\rm pr}$) and does not cause substantial difference in the posterior distribution of the initial partial pressure of \ce{N2}.}
\label{fig:post_hd}
\end{figure}

\begin{figure}[!htbp]
\centering
\includegraphics[width=1.0\textwidth]{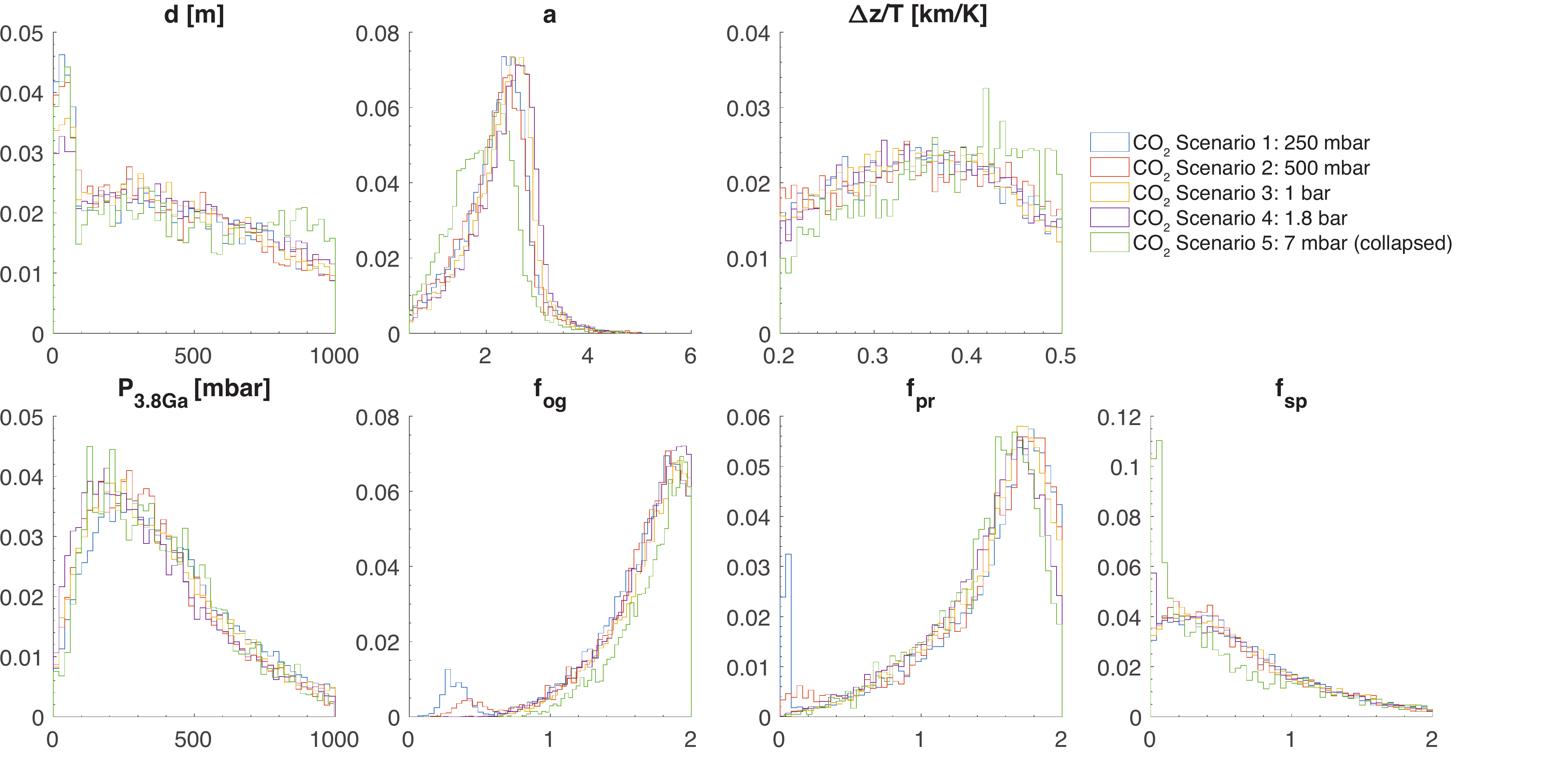}
\caption{Sensitivity to the assumed upper bound of the power-law index of photochemical loss ($a$). The simulations shown are the same as in Fig.~\ref{fig:post_2}, except that the upper bound of the power-law index is 6.0 rather than 3.0 in the standard models. The data do not require the power-law index to be greater than 3.0 in most cases, and thus the posterior distribution of the initial partial pressure of \ce{N2} is not significantly impacted by the choice of the upper bound for the power-law index.}
\label{fig:post_a}
\end{figure}

\begin{figure}[!htbp]
\centering
\includegraphics[width=1.0\textwidth]{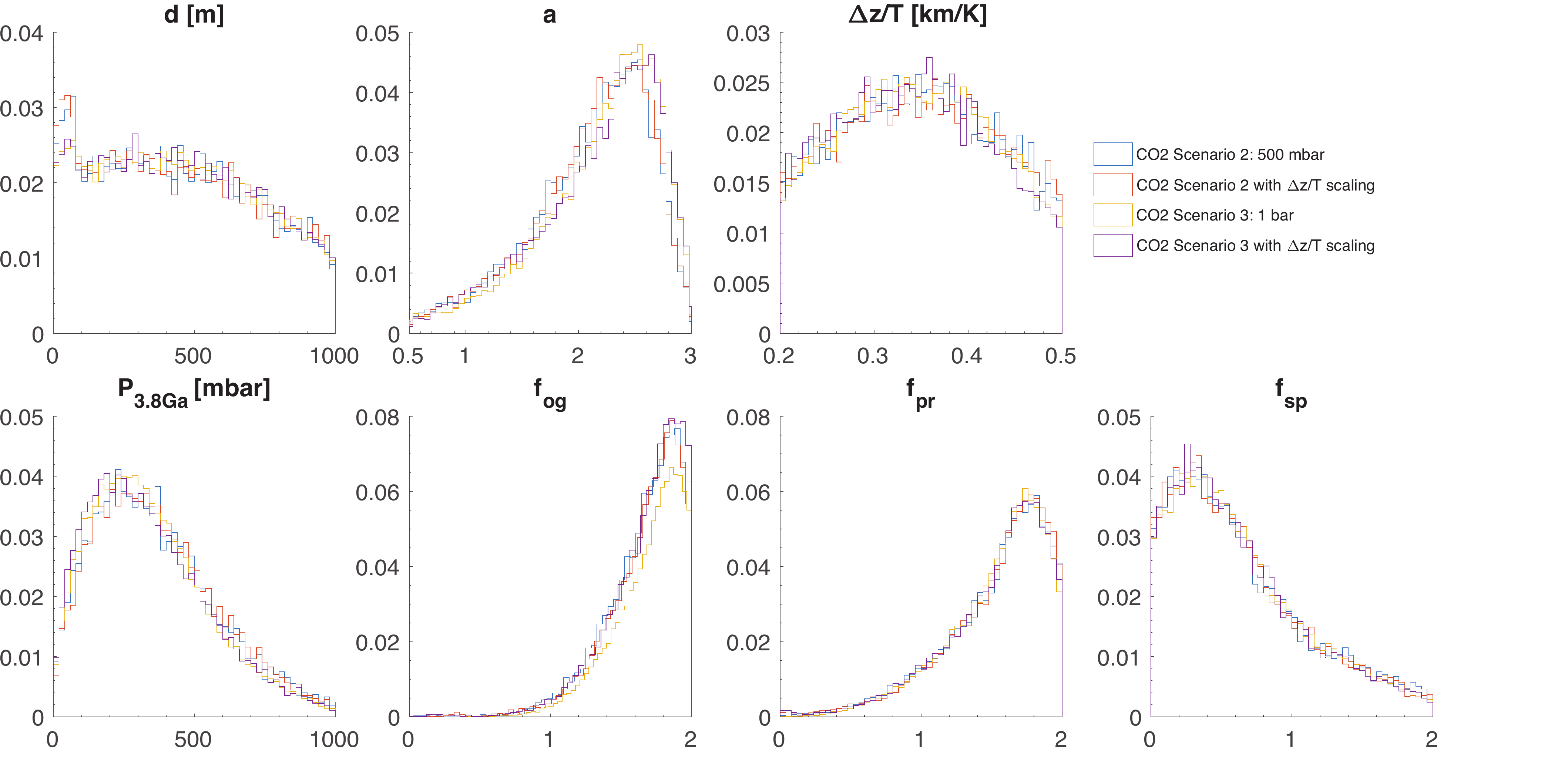}
\caption{Sensitivity to the $\Delta z/T$ scaling. For two simulations with the \ce{CO2} evolution scenarios No. 2 and 3, additional simulations are shown with the scaling of the $\Delta z/T$ parameter additionally applied. The scaling modifies $\Delta z/T$ as inverse-proportional to the mean molecular weight of the atmosphere to mimic a scale height variation for a \ce{N2}-rich atmosphere. The scaling does not cause substantial changes in the posterior distributions.}
\label{fig:post_z}
\end{figure}

We have run MCMC simulations for varied initial $\delta^{15}$N values and specific implementation of the baseline volcanism model. We have also tested sensitivity on the assumed upper bound of the power-law index of photochemical loss ($a$) and a $\Delta z/T$ scaling. The resulting posterior distributions are shown in Figs.~\ref{fig:post_hd}-\ref{fig:post_z}.

The constraints on the initial partial pressure of \ce{N2} are not sensitive to the initial $\delta^{15}$N value assumed in the simulations (Fig.~\ref{fig:post_hd}), and this is because most of the dynamical solutions are close to the transitional solutions. The sensitivity on the baseline outgassing model is largely compensated by the uncertainty in $\Delta z/T$ and only leads to moderate changes in the constraints on the initial partial pressure (Fig.~\ref{fig:post_hd}).

The wide range of the power-law index $a$ explored in this study (Table~\ref{table:priors}) captures the uncertainty in the age dependency of the solar EUV flux. Tu et al. (2015) suggested that the range of the exponent for power-law fits of the $F_{\rm EUV}$ evolution would be between $-0.96$ and $-2.15$, corresponding to the 10th and 90th percentile of the distribution of possible early solar rotation rates\cite{tu2015extreme}. Given that the saturation phase of the Sun concluded $\sim100$ Myr after formation\cite{jackson2012coronal}, the power-law fit of $F_{\rm EUV}$ is a valid approximation within our model domain. For a ``fixed'' exponent ($-1.23$) of the EUV luminosity history, the range of $a$ in 0.5 -- 3.0 corresponds to a compounded power-law exponent between $-0.62$ and $-3.96$ as a function of time. Technically, the compounded exponent may be steeper than $-3.96$ if both $a$ and the age dependency are close to their upper bounds; nonetheless, the MCMC simulations with an extended upper bound for $a$ (from 3.0 to 6.0) do not show any appreciable changes (Fig.~\ref{fig:post_a}). Our parameterization to consolidate the uncertainty into $a$ thus explores a reasonably wide range of early solar rotation rates.

It is conceivable that the value of $\Delta z/T$ may vary over the course of an evolution because it may depend on the \ce{N2}/\ce{CO2} mixing ratio. We additionally consider this effect, by scaling the value of $\Delta z/T$ as inverse-proportional to the mean molecular weight of the atmosphere to mimic the scale height variation from an \ce{N2}-poor atmosphere to an \ce{N2}-rich atmosphere. However, the MCMC simulations using the scaling do not lead to any appreciable changes in the posterior distributions from the standard models (Fig.~\ref{fig:post_z}), indicating that the potential effect of $\Delta z/T$ variation is limited.

\clearpage
\newpage

\section*{Supplementary Information G: Coupled carbon and nitrogen evolution models}

In this section, we present coupled and self-consistent carbon and nitrogen evolution models. These models are comparable to the representative cases of Fig.~\ref{fig:example}, but now include a self-consistent tracing of the \ce{CO2} abundance and $\delta^{13}$C. These models confirm that an initially large \ce{N2} reservoir can exist alongside an initially large \ce{CO2} reservoir and still satisfy the present-day constraints on the Martian atmosphere.

The \ce{CO2} evolution model is based on the model in Hu et al. (2015)\cite{Hu2015} but includes a few modifications. The most important modification is that the calculation of the escape rates of carbon now takes into account the mixing ratio of \ce{CO2}, which is no longer defaulted to unity as before. The coupled model thus proportionally reduces the escape rates of carbon when \ce{N2} is an important component of the atmosphere. Additionally, to maintain consistency with the nitrogen evolution model, several technical modifications have been included. First, to trace the instantaneous mixing ratios of \ce{CO2} in the atmosphere, the partitioning of \ce{CO2} among the atmosphere, the polar caps, and the regolith is modeled. Before, these components are treated as a combined reservoir. Here we adopt a simplifying assumption that \ce{CO2} was mostly in the atmosphere before the atmospheric collapse, and after the collapse, 7 mbar of \ce{CO2} was in the atmosphere in vapor equilibrium with the polar caps. We assume that the collapse occurred when the total pressure of the atmosphere drops below 500 mbar\cite{forget20133d}. Second, the crustal production rate profile as shown in Fig.~\ref{fig:crustal} is now adopted as the basis for calculating the outgassing rate in the \ce{CO2} evolution. Third, for completeness, we have included ion escape of \ce{CO2+} utilizing the same parametric model\cite{manning2011parametric} we have utilized for the \ce{N2+} escape in this work. In total, the updated \ce{CO2} evolution includes sputtering loss, photochemical loss, ion loss, carbonate deposition, and volcanic outgassing. The three escape processes depend on the mixing ratio of \ce{CO2} in the atmosphere, which now depends on the partial pressure of \ce{N2} in the atmosphere, making the evolution self-consistent.

\begin{figure}[!htbp]
\centering
\includegraphics[width=0.6\textwidth]{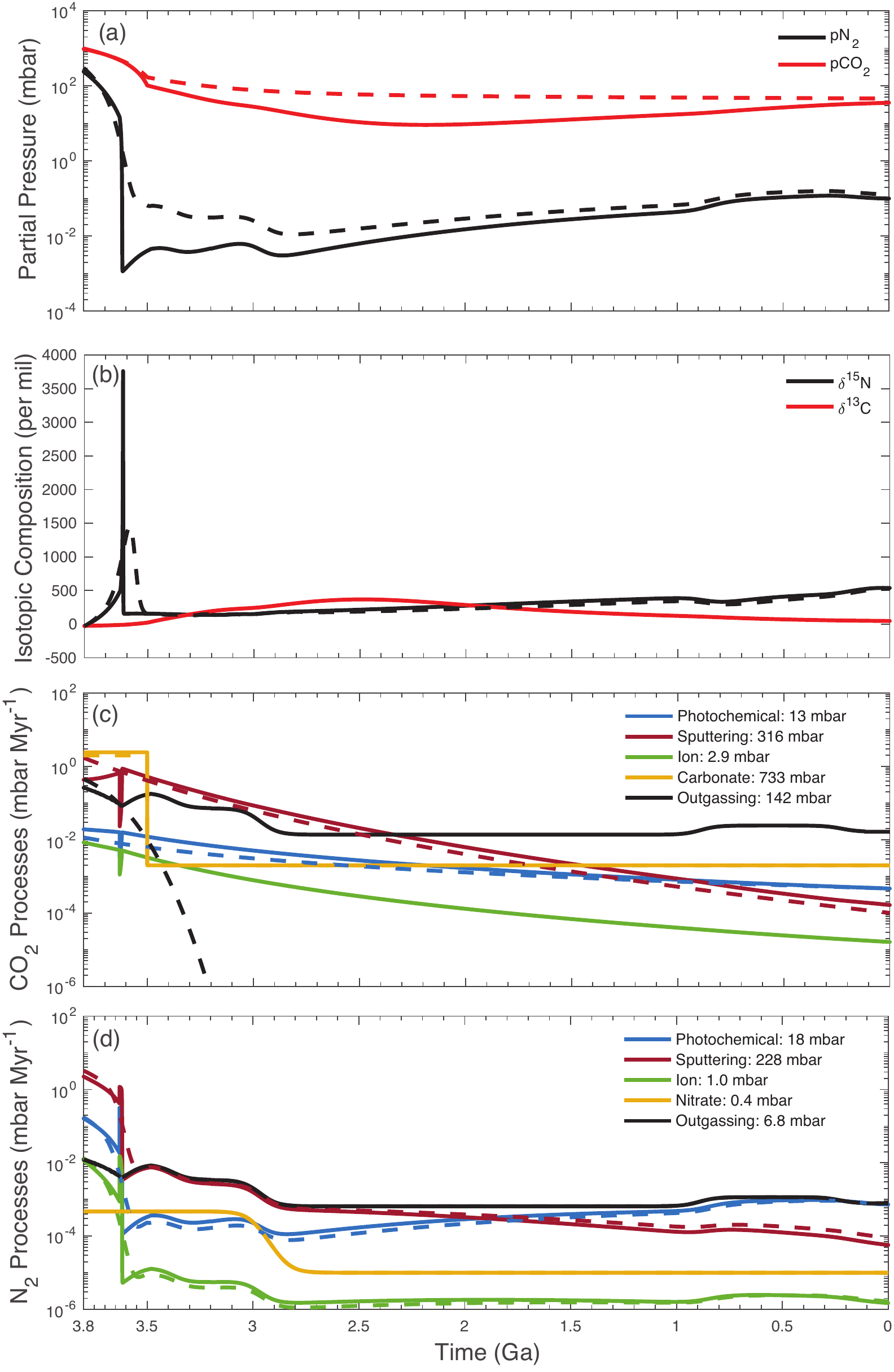}
\caption{Example of the self-consistent steady-state solutions. Dashed lines correspond to the \ce{N2} evolution shown in the left column of Fig.~\ref{fig:example} and the adopted \ce{CO2} evolutionary scenario No.3 shown in Fig.~\ref{fig:CO2}, and solid lines are the self-consistent solutions. In this plot, the partial pressure of CO$_2$ refers to the sum of the regolith, polar-cap, and atmospheric reservoirs. The atmospheric collapse occurred between 3.6 and 3.7 Ga, which caused a transient drop of the escape rates of \ce{CO2} and a spike of the escape rates of \ce{N2}. The steady state of \ce{N2} was quickly established when the atmospheric collapse occurred.}
\label{fig:ss2}
\end{figure}

\begin{figure}[!htbp]
\centering
\includegraphics[width=0.6\textwidth]{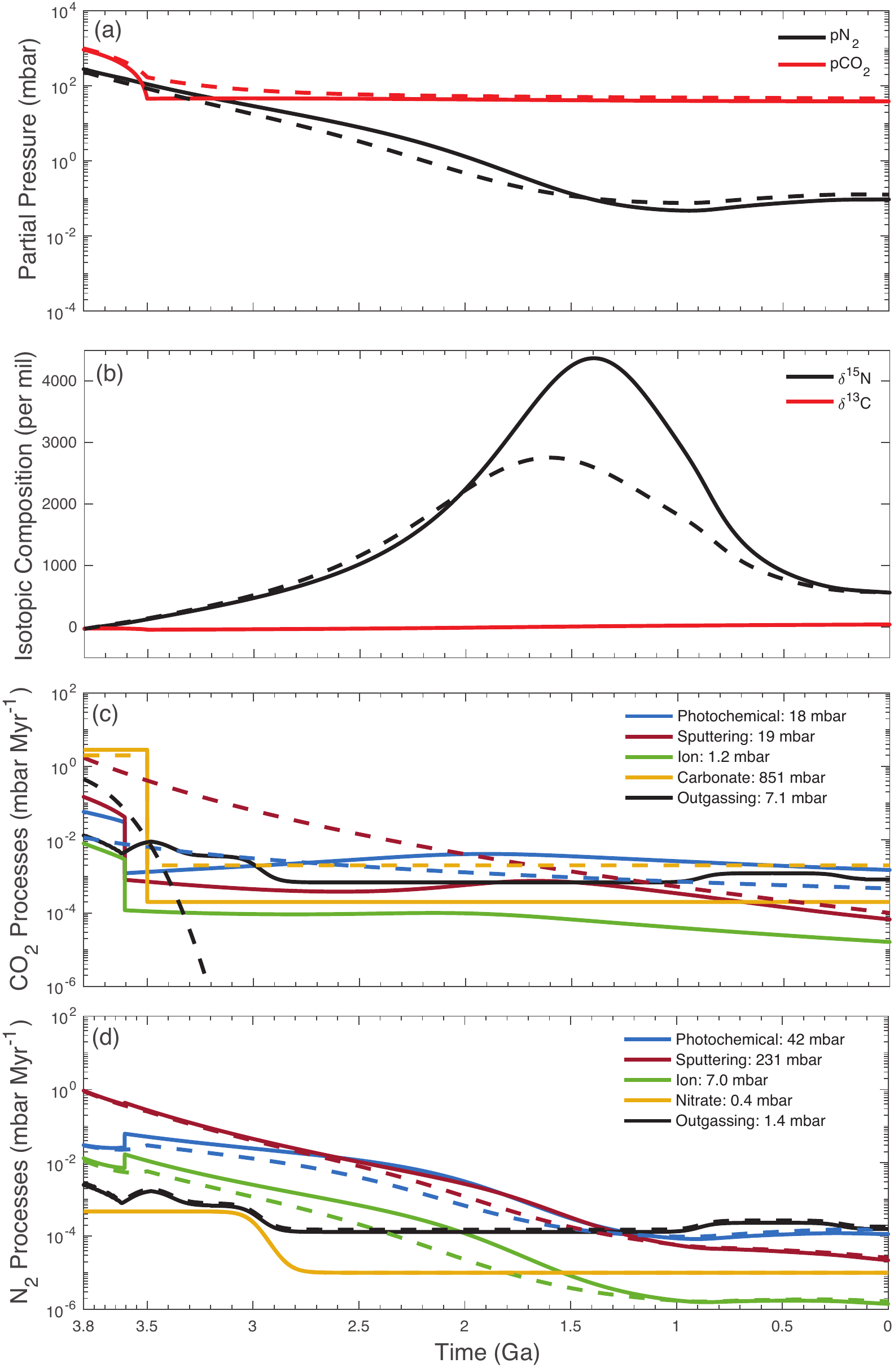}
\caption{Example of the self-consistent dynamical solutions. Dashed lines correspond to the \ce{N2} evolution shown in the right column of Fig.~\ref{fig:example} and the adopted \ce{CO2} evolutionary scenario No.3 shown in Fig.~\ref{fig:CO2}, and solid lines are the self-consistent solutions. In this plot, the partial pressure of CO$_2$ refers to the sum of the regolith, polar-cap, and atmospheric reservoirs. The atmospheric collapse occurred $\sim3.6$ Ga, which caused a long-term drop of the escape rates of \ce{CO2} and an increase of the escape rates of \ce{N2}.}
\label{fig:dyn2}
\end{figure}

We have used the coupled model to study how the inclusion of a potentially important \ce{N2} component modifies the \ce{CO2} evolution, and in turn, the \ce{N2} evolution itself. To do this, we start from the representative steady-state and dynamical solutions presented in Fig.~\ref{fig:example}. After coupling, the evolution of the carbon reservoir would result in a value of pCO$_2$ and $\delta^{13}$C different from the present day if keeping all parameters the same. As we will show in the following, however, modification of some parameters in the reasonable range can readily restore both \ce{CO2} and \ce{N2} evolutionary solutions (Figs.~\ref{fig:ss2} and \ref{fig:dyn2}).

For the steady-state solutions, the coupling generally does not impact the evolution of \ce{CO2} except for the very early period when the steady state has not been established. The model in Fig.~\ref{fig:ss2} is calculated by modifying the parameters of the steady-state solution presented as the dashed line in the left column of Fig.~\ref{fig:example}. In this model, the \ce{CO2} escape rates are reduced only when pN$_2$ is large such that it has a non-negligible effect on the mixing ratio of \ce{CO2}, which only happens before \ce{N2} descends to the steady state at $3.6-3.7$ Ga. We find that 725 mbar of early carbonate deposition (i.e., a moderate increase from 600 mbar before) would be sufficient to compensate for the reduction in the escape rates of carbon and maintain the initial \ce{CO2} partial pressure at $\sim1$ bar. Additionally, the difference in the crustal production rates results in a \ce{CO2} solution with increasing pressure during the Amazonian as opposed to the monotonically decreasing pressure of the previous \ce{CO2} solution. Then, the \ce{N2} escape rates would be moderately enhanced during the Amazonian if all parameters are kept the same. In Fig.~\ref{fig:ss2}, we have decreased the sputtering multiplier ($f_{\rm sp}$) from 3.5 to 2.5 and increased the photochemical multiplier ($f_{\rm pr}$) from 6.5 to 7 to reproduce the nitrogen solution.

The coupled evolution is typically more impacted in dynamical solutions when \ce{N2} has a non-negligible mixing ratio for a longer period. The model in Fig.~\ref{fig:dyn2} is calculated by modifying the parameters of the dynamical solution presented in the right column of Fig.~\ref{fig:example}. Here we see that the sputtering rate of \ce{CO2} is significantly reduced before $\sim1.5$ Ga, because the atmospheric pN$_2$ remains high for a long period after the atmospheric collapse, while the partial pressure of \ce{CO2} is kept at 7 mbar. We have applied 850 mbar of early carbonate deposition (compared to 600 mbar before) in this case to maintain the initial \ce{CO2} partial pressure at $\sim1$ bar. This moderate increase in the early carbonate deposition remains consistent with geologic constraints\cite{Hu2015}. 
On the nitrogen side, because slightly more sputtering loss takes place in $\sim2.5-1.5$ Ga, we have decreased the photochemical multiplier ($f_{\rm pr}$) from 1.3 to 1.15 to reproduce the present-day value of $\delta^{15}$N and decreased the outgassing multiplier ($f_{\rm og}$) from 1.5 to 1.3 accordingly to reproduce the present-day partial pressure of \ce{N2}. These moderate changes in the nitrogen evolution's parameters eventually result in a small increase of the initial pN$_2$ from 240 to 280 mbar.

To summarize, Figs.~\ref{fig:ss2} and \ref{fig:dyn2} show self-consistent, coupled \ce{CO2}-\ce{N2} solutions that reproduce the partial pressures and isotopic composition of carbon and nitrogen on present-day Mars. The character of the steady-state and dynamical evolutionary solutions remains unchanged from the stand-alone \ce{N2} models, despite the inclusion of \ce{CO2} and the modifications described above. The coupled solutions are found by moderate changes in certain input parameters based on the stand-alone solutions presented in the main text. In particular, the large initial pN$_2$ in dynamical solutions may have suppressed the escape of carbon during Hesperian and early Amazonian, and thus require a greater amount of early carbonate deposition to have an initial pCO$_2$ of $\sim1$ bar. The amount of the early carbonate deposition is however poorly constrained, and so is the initial pCO$_2$. The self-consistent solutions demonstrate that a large initial \ce{N2} reservoir is fully compatible with a large initial \ce{CO2} reservoir. \ce{N2} thus likely contributed to early surface warming on Mars.

\clearpage
\newpage

\section*{Supplementary Information H: Re-evaluation of the impact decomposition of nitrates as a potential source}

Manning et al. (2008)\cite{manning2008nitrogen} proposed an innovative scenario for the evolution of atmospheric nitrogen on Mars, in which the present-day nitrogen is in a steady state between escape and impact decomposition of previously deposited near-surface nitrates. Because we now have measurements of near-surface nitrates on Mars\cite{stern2015,sutter2017evolved}, we can re-evaluate this scenario and determine whether the impact decomposition may serve as an important source of nitrogen on today's Mars.

Based on impact models, Manning et al. (2008) estimated that the surface area affected by impact is 475 m$^2$ yr$^{-1}$ currently. Assuming an average 10-m depth affected by impacts (i.e., a generous estimate of impactors having a 10-m diameter, which only occurs once every $\sim10$ years\cite{ivanov2001mars}), and using the high end of the nitrite concentration detected on Mars (550 ppm by weight in \ce{NO3}, measured in the Cumberland rock sample by Curiosity\cite{sutter2017evolved}), we estimate an upper limit of nitrogen that can be released by impact decomposition, in planetary average, to be $\sim10^3$ N cm$^{-2}$ s$^{-1}$, or $3\times10^{-7}$ mbar Myr$^{-1}$ using the same unit for fluxes as Fig. \ref{fig:example}. The actual flux is likely lower by more than one order of magnitude because smaller impacts dominate and because the nitrite content measured in all other rock and soil samples are less than the Cumberland sample by a factor of a few. This upper-limit flux compares minimally to the present-day outgassing and photochemical escape rate of N, on the order of $10^{-4}$ mbar Myr$^{-1}$ (Fig.~\ref{fig:example}). Our estimate of the nitrate formation rate in the recent history is $10^{-5}$ mbar Myr$^{-1}$, assuming that the nitrates found in the Martian soil were formed at a constant rate during the Amazonian\cite{smith2014formation}. This rate is also significantly higher than the upper limit of nitrate decomposition. We thus find that, given the near-surface nitrate concentrations measured by Curiosity, the escape of nitrogen from Mars's atmosphere cannot be in a steady state with the impact decomposition of nitrate. 

Also, because the nitrate formation rate is modeled as a constant in this work, and because the impact decomposition flux would also be modeled as a constant, our model that explores a varied rate of nitrate formation implicitly includes the potential effect of a partial decomposition due to impact. 

\section*{Supplementary Information References}

\end{document}